\documentclass[conference]{IEEEtran}
\usepackage{fixltx2e,float}
\usepackage{graphicx}
\usepackage{epstopdf}
\usepackage{amsmath}
\usepackage{subfigure}
\usepackage{soul,url}
\usepackage{color}
\usepackage{times}
\usepackage{cite}
\usepackage[english]{babel}
\usepackage{epsfig}
\usepackage{latexsym}
\usepackage[linesnumbered,ruled,vlined]{algorithm2e}
\usepackage{caption}

\def\endofproof{\nolinebreak\ \hfill\rule{1.7mm}{1.8mm}}
\newtheorem{definition}{Definition}
\newtheorem{theorem}{Theorem}

\newcommand{\STS}{\mbox{\sf STS}\xspace}
\newcommand{\PSS}{\mbox{\sf PS$^{2}$Stream}\xspace}
\newcommand{\paratitle}[1]{\vspace{1ex}\noindent \textbf{#1}}

\begin{document}
\title{Distributed Publish/Subscribe Query Processing on the Spatio-Textual Data Stream}

\author{
{Zhida Chen$^\dag$, Gao Cong$^\dag$, Zhenjie Zhang$^\ddag$, Tom Z.J.
Fu$^\ddag$$^\sharp$, Lisi Chen$^\natural$} \vspace{1.5mm}\\
  \fontsize{11}{9} $\;^\dag$Nanyang Technological University, Singapore\vspace{0.8mm}\\
\fontsize{11}{9} $\;^\ddag$Advanced Digital Sciences Center, Illinois at Singapore Pte. Ltd., Singapore\vspace{0.8mm}\\
\fontsize{11}{9} $\;^\sharp$Guangdong University of Technology, China\vspace{0.8mm}\quad
\fontsize{11}{9} $\;^\natural$Hong Kong Baptist University, Hong Kong\vspace{0.8mm}\\
\fontsize{9}{9}\selectfont\ttfamily\upshape $\;^\dag$chen0936@e.ntu.edu.sg, gaocong@ntu.edu.sg\vspace{0.8mm}\quad
\fontsize{9}{9}\selectfont\ttfamily\upshape $\;^\ddag$zhenjie@adsc.com.sg, tom.fu@adsc.com.sg \vspace{0.8mm}\\
\fontsize{9}{9}\selectfont\ttfamily\upshape $\;^\natural$chenlisi@comp.hkbu.edu.hk
}

\maketitle
\IEEEpeerreviewmaketitle

\begin{abstract}
Huge amount of data with both space and text information, e.g., geo-tagged
tweets, is flooding on the Internet.  Such spatio-textual data stream contains
valuable information for millions of users with various interests on different
keywords and locations. Publish/subscribe systems enable efficient and
effective information distribution by allowing users to register continuous
queries with both spatial and textual constraints. However, the explosive
growth of data scale and user base has posed challenges to the
existing centralized publish/subscribe systems for spatio-textual data streams.

In this paper, we propose our distributed publish/subscribe system, called
\emph{PS$^2$Stream}, which digests a massive spatio-textual data stream and
directs the stream to target users with registered interests. Compared with
existing systems, \emph{PS$^2$Stream} achieves a better workload distribution
in terms of both minimizing the total amount of workload and balancing the
load of workers. To achieve this, we propose a new workload distribution
algorithm considering both space and text properties of the data.
Additionally, \emph{PS$^2$Stream} supports dynamic load adjustments to adapt
to the change of the workload, which makes \emph{PS$^2$Stream} adaptive.
Extensive empirical evaluation, on commercial cloud computing platform with
real data, validates the superiority of our system design and advantages of
our techniques on system performance improvement.
\end{abstract}

\section{Introduction}
\label{sec:introduction}
Driven by the proliferation of GPS-equipped mobile devices and social media
services, data containing both space and text information is being generated
at an unprecedented speed. Millions of social-media users, for example, are
uploading photos to Instagram with both location and text tags, posting
geo-tagged tweets on Twitter, and creating location-aware events in Facebook,
  using their smart phones.  Such user generated content arrives in a
  streaming manner, and contains valuable information to both individual and
  business users. On the one hand, individual users may be interested in
  events in particular regions, and are keen to receive up-to-date messages
  and photos that originate in the interested regions and are relevant to the
  events. On the other hand, business users, e.g., Internet advertisers,
  expect to identify potential customers with certain interest at a particular
  location, based on their spatio-textual messages, e.g., restaurant diners in
  a target zone. For both types of users, it is important to find
  spatio-textual messages satisfying location and textual constraints in
  real-time, to deliver the service for their business models.

Publish/subscribe systems \cite{cao2004efficient,tariq2009providing,papaemmanouil06,baldoni2005content,carzaniga2001design,yu2012subscriber} provide the basic
primitives to support such information processing paradigms, such that
\emph{subscribers} register subscription (continuous) queries in the system to
catch all messages from \emph{publishers} satisfying the query predicates. In
the context of our problem setting, each registered query from the subscribers
consists of two components, a space component describing the spatial region of
interests and a text component containing a boolean expression of keywords for
matching. A message matches a query, if its location lies in the interested
region and its textual content satisfies the boolean expression of the query.
When a massive spatio-textual data stream floods on the Internet, the
publish/subscribe system filters the messages and routes the matching messages
to subscribers in real-time.

Existing  publish/subscribe systems are capable of handling subscriptions on a
spatio-textual data stream at moderate rates, by utilizing indexing techniques
tailored for spatio-textual data on a centralized server~\cite{li2013location,
  icde15huiqi,icde15xiang,chen2013efficient,wang2016skype,yu2015cost,icde15lisi}.
  With the growth of spatio-textual messages on social media and
  registered queries, the computation workload for publish/subscribe systems
  is quickly increasing, which exceeds the capacity of a single server. This
  calls for a distributed solution to building a publish/subscribe system
  over the spatio-textual data stream.  Moreover, due to the dynamic nature of the
  social media, the workload of publish/subscribe systems varies dramatically
  over time. The system is expected to conduct dynamic load adjustments with
  small migration cost to fit for the change in the workload.

In this paper, we present our system, called \PSS, as a scalable distributed
system over a spatio-textual data stream, to enable real-time
publish/subscribe services on top of a cluster of servers.  The system design
objective is to accomplish a maximal processing throughput, minimal latency, and
ignorable migration cost. To accomplish these goals, we propose a handful of
new techniques. Firstly, we propose a new workload partitioning strategy,
    utilizing both text and space properties of the data, to facilitate
    distributing the workload with aims of minimizing the total amount of
    workload and load balancing of workers. Different from existing
    distributed systems, we need to distribute both spatio-textual data stream
    and subscription query stream, and different ways of workload distribution
    will result in different amounts of workload.  Secondly, we discuss dynamic
    load adjustment approaches over our system architecture, in both local and
    global manners, to minimize both the overhead of workload reassignment and
    the total amount of workload in the scenario of workload changing. 
   The main contributions of our work are summarized as follows:

1) We propose a new workload partitioning algorithm that considers both space
and text properties of the data for the workload distribution.
Our system is the first to consider the optimal workload
partitioning problem for distributing both spatio-textual data stream and
subscription query stream, which aims at minimizing the total amount of workload,
and balancing the load of workers. 
  
2) We propose efficient dynamic load adjustment algorithms to adjust the load
of workers in the scenario of workload changing. By considering both space
and text properties of the data, our dynamic load adjustment algorithms
achieve features of reducing the total amount of workload and invoking small
migration cost. To the best of our knowledge, our system is the first to
support such dynamic load adjustments.

3) We conduct extensive experiments on Amazon EC2 with a real-life
spatio-textual data stream. The results demonstrate that: i) our distribution
framework considering both space and text properties performs better than the
baselines utilizing either space or text property only, and our system
achieves excellent performance on both processing throughput and query
response latency; ii) our dynamic load adjustment approaches improve the
performance of the system and invoke small migration cost.

\section{Related Work}
\label{sec:related_work}

\paratitle{Spatial-Keyword Publish/Subscribe.} The problem of building
publish/subscribe systems over a spatio-textual data stream on a centralized
server has been studied recently~\cite{li2013location,
  icde15huiqi,icde15xiang,chen2013efficient,wang2016skype,yu2015cost,icde15lisi}.
  They focus on developing new indexes to speed up the matching between
  spatio-textual objects and the spatial-keyword continuous queries.
  Specifically, Chen \textit{et al}~\cite{chen2013efficient} present an
  indexing structure called IQ-tree.  Li \textit{et al}~\cite{li2013location}
  propose an R-tree based indexing structure.  Wang \textit{et
    al}~\cite{icde15xiang} propose to adaptively group the continuous queries
    using keyword partitions or space partitions based on a cost model.  The
    problem of similarity based spatial-keyword publish/subscribe is also
    studied~\cite{icde15huiqi,icde15lisi}, which determines the matching
    between a subscription and a published spatio-textual object based on a
    similarity score.  

In theory, these studies are complementary to our work as these
complicated index structures could be employed in workers of our system.
However, it would be expensive to jointly maintain and migrate these index
structures across multiple workers, which is required in our system. 

\paratitle{Distributed Content based Publish/Subscribe.} Our work is related
to content based subscription queries.  
Many works~\cite{cao2004efficient,tariq2009providing,papaemmanouil06,baldoni2005content,carzaniga2001design,yu2012subscriber}
exist for distributed content-based publish/subscribe systems over a wide-area
network. 
They aim at finding an optimal assignment of the subscribers to message
brokers with some performance criterion such as the
latency~\cite{tariq2009providing}. 
Another line of works\cite{terpstra03,meghdoot} consider deploying the
publish/subscribe systems on a P2P network, which focus on minimizing the
communication cost. E-StreamHub~\cite{barazzutti2014elastic} is a distributed
content based publish/subscribe system deployed on a cluster of local servers.
However, these systems do not provide supports for the spatio-textual data,
  which is essential to our problem, and we do not see any sensible way to
  deploy these systems for our problem.

Our system differs from these systems in two aspects. Firstly, our system is based
on the optimal workload partitioning problem, which aims at both of minimizing
the total amount of workload and balancing the load of workers in distributing
the workload of both subscription query stream and spatio-textual data stream. Secondly, we
support dynamic load adjustments by considering minimizing both total amount
of workload and migration cost. 

\paratitle{Distributed Systems for Spatial Data.} Many distributed
systems~\cite{aji2013hadoop, eldawy2015spatialhadoop, akdogan2010voronoi,
  nishimura2011md, aly2015aqwa} have been proposed for
  large-scale spatial data.  
Most of them use existing spatial indexes to partition the data to different
servers, e.g., SpatialHadoop~\cite{eldawy2015spatialhadoop} uses a grid index
and a R-tree, and MD-HBase~\cite{nishimura2011md} uses a kd-tree and a Quad
tree.  AQWA~\cite{aly2015aqwa} uses a kd-tree to partition the spatial data,
  which aims at minimizing the querying cost based on a query workload. Our
  system differs from these systems in two ways. Firstly, we consider the
  workload composed of updating highly dynamic subscription queries and
  processing spatio-textual objects, while they consider processing disposable
  queries on a static spatial data set. Secondly, our system design objectives
  are different. We consider minimizing the total amount of workload and load
  balancing of workers in distributing workload, as well as dynamic load
  adjustments. We evaluate for the first time using the data partitioning
  strategies in these systems for our problem in our experiments.
  
\paratitle{Distributed Data Stream Processing.} The distributed data stream
processing systems are also related to our work.  A common scheme is to split
a continuous query into multiple operators and distribute them to a cluster of
servers. 
NiagaraCQ~\cite{chen2000niagaracq} detects the commonality of different
continuous queries and combines their common operators.
PSoup~\cite{chandrasekaran2003psoup} improves the processing efficiency of
stateful operators such as join and aggregate. 
Subsequent systems like Aurora*\cite{cherniack2003scalable} and
Borealis~\cite{abadi2005design} further improve the scalability of the data
stream processing systems by allowing the same operator being executed by
multiple servers. These systems lack support for processing the spatio-textual
data stream and their proposed methods do not apply for our system.

A recent demonstration system, called Tornado~\cite{mahmood2015tornado}, is
presented for indexing spatio-textual data streams to handle querying
spatio-textual data streams. However, it does not handle the stream of
subscription queries. Furthermore, unlike our system, Tornado does not
consider the optimal workload partitioning problem, and it does not support
dynamic load adjustments which reduce the total amount of workload with small
migration cost. Tornado uses a kd-tree to partition the workload, which has
been included in our evaluation of the baselines.

\section{Problem Statement and System Architecture}
\label{sec:prob_system}

\subsection{Problem Statement}
\label{sec:problem_statement}

\noindent \textbf{Spatio-Textual Object:} A spatio-textual object is defined
as $o=\langle text, loc\rangle$, where $o.text$ is the textual content of
object $o$ and $o.loc$ is the geographical coordinate, i.e., latitude and
longitude, of object $o$.

In this work, we consider a stream of spatio-textual objects, such as
geo-tagged tweets, check-ins in Foursquare, geo-tagged posts in Instagram,
etc.  We aim at building a distributed publish/subscribe system over a stream
of spatio-textual objects.

Users may express their interests on the spatio-textual objects with
subscription queries. Following the previous works~\cite{li2013location,
icde15xiang,chen2013efficient}, each subscription query contains a Boolean
keyword expression and a region. If a new spatio-textual object falls in the
specified region and satisfies the Boolean keyword expression specified by a
subscription query, the object will be pushed to the user who submits the
query. A subscription query is valid until the user drops it. We next present
the spatio-textual subscription query.

\smallskip \noindent \textbf{Spatio-Textual Subscription (STS) Query:} A
Spatio-Textual Subscription (\STS) Query is defined as $q=\langle K,R\rangle$,
where $q.K$ is a set of query keywords connected by AND or OR operators, and
$q.R$ denotes a rectangle region.

A spatio-textual object $o$ is a result of an \STS query $q$ if $o.text$
satisfies the boolean expression of $q.K$ and $o.loc$ locates inside $q.R$.

The large number of \STS queries and high arrival rate of spatio-textual
objects call for a distributed solution. We build our system on a cluster of
servers with several servers playing the role of \textit{dispatchers}, which
distribute the workload to other servers. The workload to our system includes
the insertions and deletions of \STS queries, and the matching operations
between \STS queries and spatio-textual objects. 

\smallskip \noindent {\bf Problem Statement} We aim at building a distributed
publish/subscribe system over a spatio-textual data stream. We expect the system to
have the features that (i) the system can achieve a high throughput, (ii) each
tuple can be processed in real-time, i.e., low latency, and (iii) the system
can dynamically adjust the load of servers when the workload changes.

\subsection{System Architecture}
\label{sec:system_architecture}

\begin{figure}[!ht]
\vspace{-3ex}
\begin{minipage}[c]{\linewidth}
\center
\includegraphics[scale=0.42]{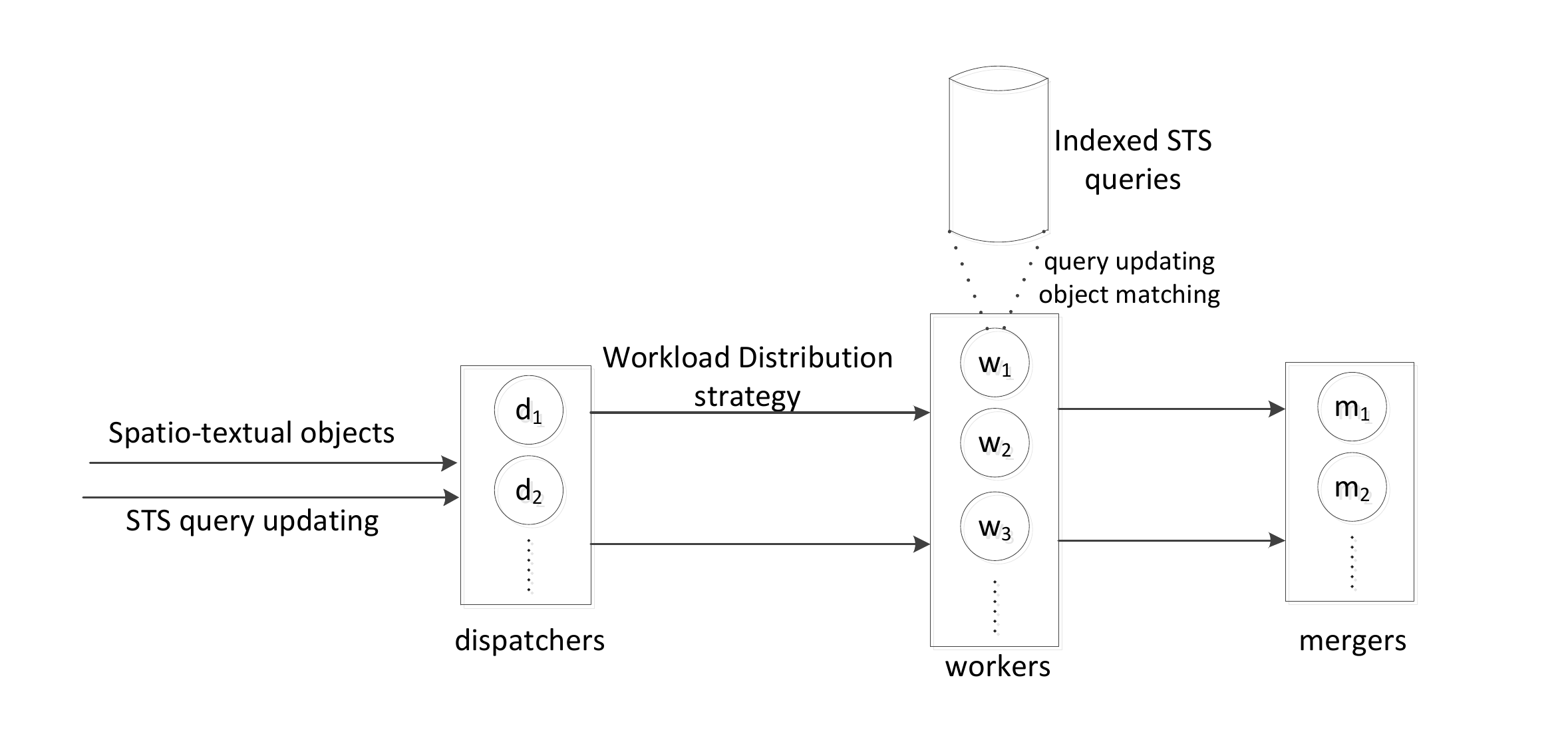}
\end{minipage}
\vspace{-2ex}
\caption{Architecture of PS$^2$Stream}
\vspace{-3ex}
\label{fig:fig2}
\end{figure}

We present a distributed \underline{P}ublish/\underline{S}ubscribe system for
handling subscription queries over a \underline{S}patio-textual data
\underline{stream}, and we call the system \PSS.  There are three components
in \PSS: dispatcher, worker and merger, which are shown in
Figure~\ref{fig:fig2}. The dispatcher takes as input a stream of
spatio-textual objects, and it receives two types of requests from users,
namely submitting new subscriptions and dropping existing subscriptions, which
correspond to two operations, i.e., query insertions and deletions,
respectively.  The dispatcher assumes the responsibility of distributing
spatio-textual objects and query insertions/deletions. The worker accepts the
workload sent from the dispatcher and conducts the following operations
accordingly:

\vspace{-1.0mm}\noindent
(1) \emph{Query insertion}: On receiving a new \STS query, the worker inserts it into an in-memory index maintained in the worker;

\vspace{-1.0mm}\noindent
(2) \emph{Query deletion}: On receiving the indication of deleting an existing \STS query, the worker removes the query from the index;

\vspace{-1.0mm}\noindent
(3) \emph{Matching a spatio-textual object}: On receiving a spatio-textual
object, the worker checks whether the object can be a match for any \STS query
stored in the worker. If yes, the matching result is forwarded to the merger.

\noindent
The merger assumes the responsibilities of removing duplicated matching
results produced by workers and sending the results to
corresponding users.

\section{Hybrid Partitioning}
\label{sec:hybrid}

In this section, we present the design of our \PSS system. We define a new
workload distribution problem in Section~\ref{sec:definition}, and provide a
solution to it in Section~\ref{sec:workload_part}. We introduce the
index structure adopted in dispatchers in Section~\ref{sec:index}. Finally, we
discuss how workers process their workload in
Section~\ref{sec:index_structure_on_workers}.

\subsection{Definition}
\label{sec:definition}

We propose \PSS for processing a stream of subscription queries over a
spatio-textual data stream. Different from most existing distributed systems,
we need to distribute the workload of both updating highly dynamic \STS
queries and processing spatio-textual objects. In our problem, different ways
of workload distribution will result in different amounts of workload.  During
the workload distribution, the following two factors are important for the
system.  One is the total amount of workload distributed to the workers, and
the other one is the load balancing of workers.

Next, we first define the load of one worker. Based on it, we propose our
Optimal Workload Partitioning problem. To the best of our knowledge, our
definition of the workload distribution problem is the first for distributing
spatio-textual data that considers minimizing the total amount of workload
while balancing the load of workers.

\begin{figure}
\begin{minipage}[c]{\linewidth}
\centering
\includegraphics[scale=0.5]{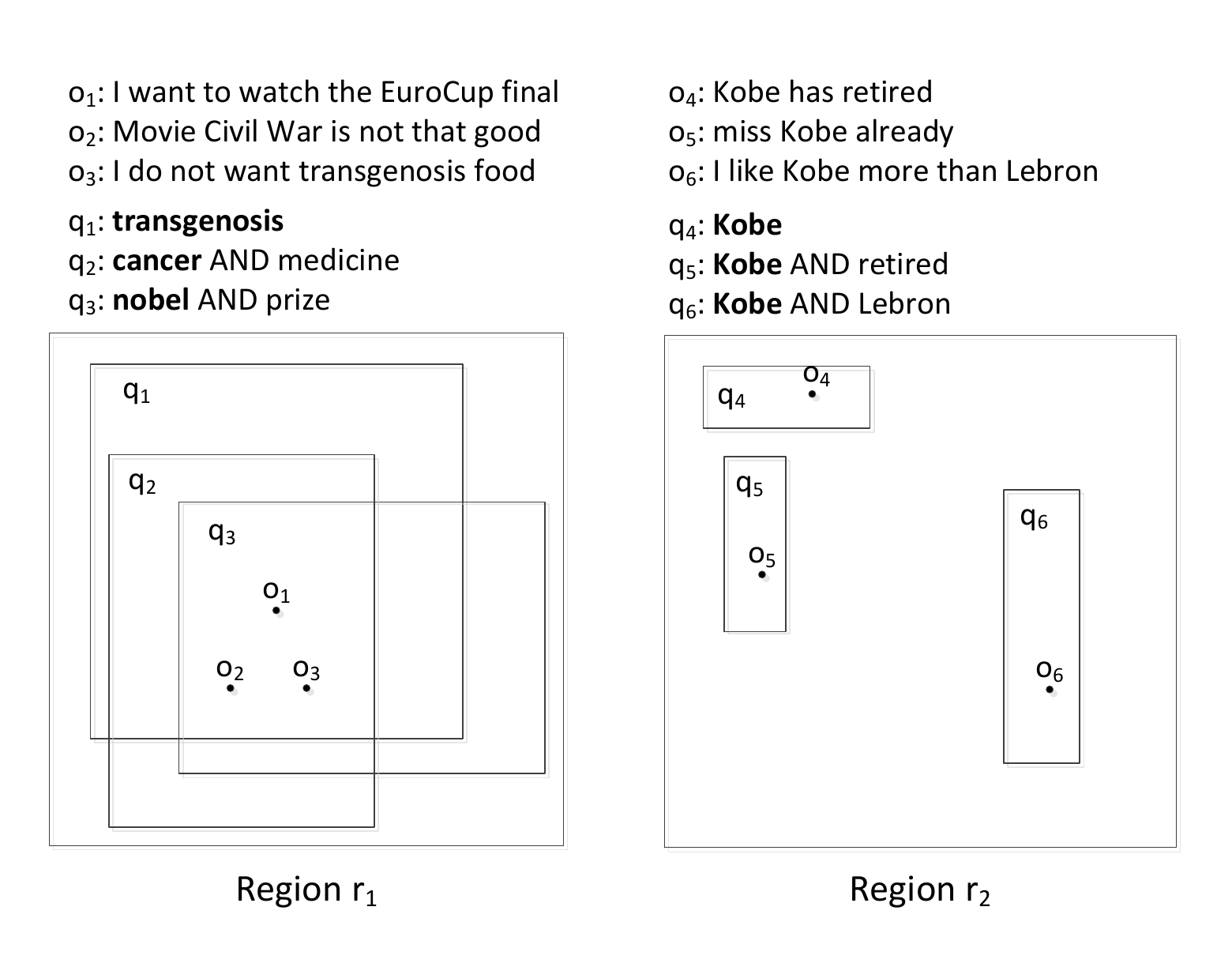}
\end{minipage}
\vspace{-2ex}
\caption{Two regions with different data distributions.}
\label{fig:diff-data-dist}
\vspace{-5ex}
\end{figure}

\begin{definition}
  \label{def:worker_load} \textbf{Load of One Worker}:
  Given a time period, the load of a worker $w_i$ for the period is computed by
  \[L_i = c_1 \cdot |O_i| \cdot |Q_{i}^{i}| + c_2 \cdot |O_i| + c_3 \cdot
  |Q_{i}^{i}| + c_4 \cdot |Q_{i}^{d}|,\]
  where $O_{i}$ is the set of
  spatio-textual objects sent to $w_i$, $Q_{i}^{i}$ is the set of \STS
  query insertion requests sent to $w_i$, and $Q_{i}^{d}$ is the set of \STS
  query deletion requests sent to $w_i$ in the given time period.
  Here $c_1$ is the average cost of
  checking whether a spatio-textual object matches an \STS query, $c_2$ is the
  average cost of handling one object, $c_3$ is the average cost of handling
  one \STS query insertion request, and $c_4$ is the average cost of handling
  one \STS query deletion request. \endofproof
\end{definition}

\begin{definition}
  \label{def:opt_work_part}
\textbf{Optimal Workload Partitioning Problem}:
Given a set of spatio-textual objects $O$, a set of \STS query insertion
requests $Q^i$ and a set of \STS query deletion requests $Q^d$, let $S$ denote
the space they reside in and $T$ denote the set of terms appearing in them.
We aim to partition $S$ into $m$
subspaces $S_1,S_2,\cdots,S_m$ and partition $T$ into $m$ subsets
$T_1,T_2,\cdots,T_m$, and assign each pair $(S_i, T_i)$ to one worker, where
$m$ is the number of workers. An object $o$ is sent to worker $w_i$ only if
$o.loc\in S_i$ and $o.text \cap T_i\neq \emptyset$. An insertion/deletion
request of an \STS query $q$ is sent to worker $w_i$ only if $q.R\cap S_i\neq
\emptyset$ and $q.K \cap T_i\neq \emptyset$. The Optimal Workload Partitioning
problem is to find a partition such that
$\sum_{i=1}^{m}L_i$ is minimized, subject to the constraint that
$\forall{i\neq j}$ and $L_i \geq L_j ,L_i / L_j
\leq \sigma$, where $\sigma$ is a small constant value being larger than 1.
\endofproof
\end{definition}

Note that the constraint in the problem statement is to achieve load balancing.
\begin{theorem}
  \label{th:opt_workload}
   The Optimal Workload Partitioning problem is NP-hard.
\end{theorem}
\noindent\textbf{Proof}: We prove by reducing from the Partition
problem, which is NP-hard. In the Partition problem, given $n$ numbers
$a_1, a_2, \cdots , a_n\in \mathbf{N}$, we decide whether there
exists a set $S\subseteq \{1, 2, \cdots , n\}$ such that $\sum_{i\in S}a_i =
\sum_{i\notin S}a_i$. For any instance of the Partition problem, we can reduce
it to one instance of the Optimal Workload Partitioning Problem in the
following way:

First, we set the value of $m$ to 2 and set the value of $\sigma$ to 1, which
means that there are 2 workers in the system and we want them to be perfectly
load balancing.

Second, we produce the workload of processing the spatio-textual objects only
($Q^i=\emptyset$ and $Q^d=\emptyset$). The objects share the same textual
content and are distributed to $n$ non-overlapping regions
$r_1,r_2,\cdots,r_n$. Let $O_i$ denote the set of objects locating in region
$r_i$ and $|O_i|=a_i$.  

Since all objects share the same textual content, to satisfy the load
balancing constraint, we cannot partition them by the text property, i.e., the
complete term set $T$ is assigned to all workers. Hence the solution is to
find a 2-partition $R_1,R_2$ of $\{r_1,r_2,\cdots,r_n\}$ satisfying that $\sum_{r_i\in
R_1}|O_i|=\sum_{r_j\in R_2}|O_j|$. It is equivalent to the Partition problem,
thus completing our proof.
\endofproof

\subsection{Workload Partitioning}
\label{sec:workload_part}

Due to the hardness of the Optimal Workload Partitioning problem, we
investigate heuristic algorithms to partition the workload. The text property
and the space property of the data inspire us to partition the workload by the
text property or the space property. Previous works~\cite{cambazoglu2013term,
basik2015s} on constructing a distributed text retrieval system or a
distributed information dissemination system use text-partitioning to
partition the workload. However, they focus on the communication cost and load
balancing.
Other works on constructing a distributed system for a static
spatial data set~\cite{aji2013hadoop, eldawy2015spatialhadoop, akdogan2010voronoi,
  nishimura2011md, aly2015aqwa} and for a spatio-textual data
stream~\cite{mahmood2015tornado}, use space-partitioning to partition the
workload, and they do not consider minimizing the total amount of workload and
the load balancing collectively.

Such partitioning algorithms based on an unanimous scheme perform poorly in
minimizing the total amount of workload when the data distributions among
different regions are quite different. As shown in
Figure~\ref{fig:diff-data-dist}, space-partitioning performs well in region
$r_2$ where the objects and queries are well spread, but it does not fit for
region $r_1$ as the ranges of queries in $r_1$ are large and clustered.
Similarly, text-partitioning is good for $r_1$ but bad for $r_2$.

To better solve the Optimal Workload Partitioning problem, we propose a new
partitioning algorithm that utilizes both text and space properties
of the data. Our hybrid partitioning algorithm decomposes the workload into a
set of units by wisely using the text property or the space property in
different regions, with the goal of minimizing the total amount of workload
after the decomposition. Additionally, we distribute those units to workers,
so that the load balancing constraint can be satisfied.

\paratitle{Algorithm overview.} The core idea of our algorithm is to first use
space-partitioning to identify the subspaces that have large differences in
the text distribution between objects and queries, and then for each subspace,
we choose between space-partitioning and text-partitioning to minimize the
total amount of workload. 

The algorithm can be divided into two phases. In the first phase, the
algorithm divides the space into subspaces based on the text similarity
between objects and queries. The purpose is to identify those subspaces where
the text-partitioning would perform better. To achieve this, for a subspace
where the text similarity between objects and queries is smaller than a
threshold, we check whether partitioning it will result in new subspace(s)
where the text similarity becomes smaller. If yes, we partition the subspace
recursively and perform the same checking for each new subspace. At the end of
this phase, we obtain two types of subspaces represented by $N_s$ and $N_t$,
respectively.  The subspaces in $N_t$ have small text similarity between
objects and queries, and we partition them using text-partitioning only in the
second phase. For the subspaces in $N_s$, we compare the workloads produced by
space-partitioning and text-partitioning, respectively, and select the
strategy resulting in a smaller workload.  This comparison is necessary as
when the query ranges are very large, using space-partitioning will invoke
queries being duplicated to multiple workers, which may result in a larger
workload than using text-partitioning even though the objects and queries have
large text similarity. In the second phase, if the number of partitions is
smaller than the number of workers, we further partition the workload in $N_s$
and $N_t$.  To minimize the total amount of workload, we design a dynamic
programming function to find the optimal number of partitions for each
subspace in $N_s$ and $N_t$. We then check whether the load balancing
constraint can be satisfied. If no, we recursively further partition one
subspace until the load balancing constraint can be satisfied.

\begin{figure}
\begin{minipage}[c]{\linewidth}
\center
\includegraphics[scale=0.5]{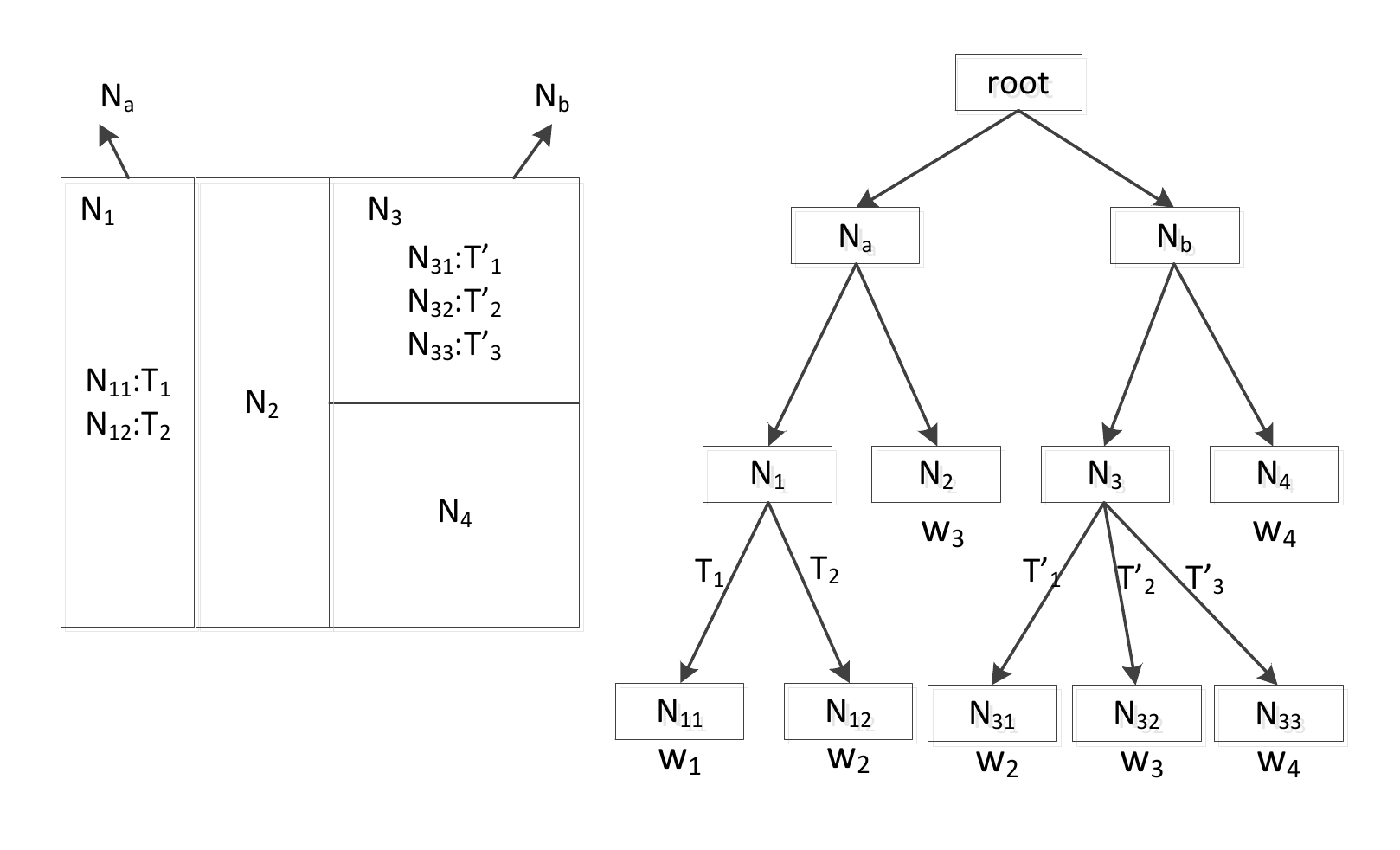}
\end{minipage}
\vspace{-2ex}
\caption{Example of a kd$^t$-tree}
\vspace{-4ex}
\label{fig:kd-t}
\end{figure}

The output of our workload partitioning procedure is an index structure named
kd$^t$-tree, which is a kd-tree with some leaf nodes being further partitioned
by the text property. 
Figure~\ref{fig:kd-t} shows an example
of a kd$^t$-tree, which distributes the workload to 4 workers: the set of leaf
nodes is divided into 4 disjoint subsets and each subset is assigned to one
worker randomly, e.g., $\{N_{11}\}$ is assigned to worker $w_1$ and
$\{N_{12},N_{31}\}$ is assigned to worker $w_2$.  For an object and query
locating inside or overlapping with the space range of node $N_2$ or $N_4$, it
is sent to worker $w_3$ or $w_4$ without checking the textual content. When an
object or query locates inside or overlaps with the space range of node $N_1$,
its textual content is checked and it is sent to worker $w_1$ (resp.  $w_2$)
if it contains terms in $T_1$ (resp. $T_2$).  Workers $w_2$, $w_3$, and $w_4$
also receive objects and queries that locate inside or overlap with the space
range of node $N_3$ and contain terms in $T_{1}^{'}$, $T_{2}^{'}$ and
$T_{3}^{'}$, respectively.

\begin{algorithm}[!ht]
  \small
  \caption{\bf{Partition Workload}}
  \label{alg:part_work} 
  \KwIn{The number of workers $m$, Load balance constraint threshold
    $\sigma$, Text similarity
    threshold $\delta$, A set of objects $O$, A set of queries $Q$;}
    \KwOut{The root node of a kd$^t$-tree with each leaf node assigned a number in
      $\{1, 2, \cdots,m\}$}

  \SetKwFunction{computenumberpartitions}{ComputeNumberPartitions}
  \SetKwFunction{partitionnode}{PartitionNode}
  \SetKwFunction{mergenodesintopartitions}{MergeNodesIntoPartitions}

  Initialize the kd$^t$-tree structure with a root node $n_r(O,Q)$\;
  $N_u \leftarrow \{n_r\}$, $N_t \leftarrow \emptyset$, $N_s \leftarrow \emptyset$\;
  \While{$|N_u|>0$}
  {
    Pop one node $n$ from $N_u$\;
    \If {$sim_t(O_{n}, Q_{n}) \geq \delta$}
    {
      $N_s \leftarrow N_s \cup \{n\}$\;
    }
    \Else
    {
      $\{n_1,n_2\} \leftarrow$ partition $n$ into 2 nodes in the direction that
       minimizes $\alpha=\textrm{min}\{sim_t(O_{n_1}, Q_{n_1}), sim_t(O_{n_2}, Q_{n_2})\}$\;

      \If {$|\alpha-sim_t(O_{n}, Q_{n})|\approx 0$}
      {
        $N_t \leftarrow N_t \cup \{n\}$\;
      }
      \Else
      {
        $N_u \leftarrow N_u \cup \{n_1, n_2\}$\;
      }
    }
  }
  \If{$|N_t|+|N_s|<m$}
  {
    $A\leftarrow \computenumberpartitions(N_t,N_s,m)$\;
    \For{each node $n$ in $N_t \cup N_s$}
    {
      \partitionnode($n$, $N_t$, $N_s$, $A[n]$)\;
    }
  }
  \While{True}
  {
    \mergenodesintopartitions($N_t$, $N_s$, $m$)\; 
    $L_{max}\leftarrow$ the maximum load value among all partitions\;
    $L_{min}\leftarrow$ the minimum load value among all partitions\;
    \If{$L_{max}/L_{min}\leq \sigma$}
    {
      break\;
    }
    \Else
    {
      $n\leftarrow$ the node having the largest load value in $N_t \cup N_s$\;
      \partitionnode($n$, $N_t$, $N_s$, 2)\;
      \If{$|N_t| + |N_s|\geq\theta$} 
      {
        break\;
      }
    }
  }

  return $n_r$\;
\end{algorithm}

Algorithm~\ref{alg:part_work} presents the pseudo code of our workload
partitioning algorithm. The algorithm first initializes a root node $n_r(O,Q)$
of the kd$^t$-tree and puts it into $N_u$ (lines 1--2). The variable $N_u$
denotes the set of nodes that have not been decided to put into $N_t$ or
$N_s$.  A while loop is conducted to compute $N_t$ and $N_s$ (lines 3--12). At
each iteration, it pops one node $n$ from $N_u$ (line 4) and computes the text
similarity between the objects and queries in $n$. We use cosine similarity in
our algorithm. If the text similarity is larger than a threshold $\delta$, we
consider node $n$ as being not suitable for text-partitioning, and add it into $N_s$
(line 6).  Otherwise, we split node $n$ in either x-dimension or y-dimension
as the normal kd-tree does, except that we prefer the direction resulting in a
smaller text similarity $\alpha$ between objects and queries in the new
subspaces (line 8). If the difference between $sim_t(O_n , Q_n)$ and $\alpha$
is minor, we consider node $n$ as being consistent in text similarity of the
objects and queries in it and add $n$ into $N_t$ (lines 9--10).  Otherwise, we
add the new nodes $n_1$ and $n_2$ into $N_u$ (line 12).  In the second phase,
the algorithm proceeds to check the number of nodes in $N_t$ and $N_s$ (line
13). If the number of nodes is smaller than $m$, i.e., the number of workers,
a dynamic programming function \textsf{ComputeNumberPartitions} is called to
compute the number of partitions for each node (lines 13--14). It is for
minimizing the total amount of workload by determining an optimal number of
partitions for each node.  The function \textsf{PartitionNode} is then called
to partition the nodes (lines 15--16).  After that, the algorithm recursively
checks whether the load balancing constraint can be satisfied (lines 18--22)
and partitions the node having the largest load, until the load balancing
constraint is satisfied or the number of nodes reaches a threshold $\theta$,
where $\theta$ is a threshold of the maximum number of nodes (lines 24--27).
In the end, the algorithm returns the root node of the kd$^t$-tree (line 28).

\paratitle{Computing the number of partitions.}
For the purpose of minimizing the total amount of workload, a dynamic programming function
\textsf{ComputeNumberPartitions} is called to compute the number of partitions
for each node in $N_t$ and $N_s$ when $|N_t|+|N_s|<m$ (lines 13--14 of
Algorithm~\ref{alg:part_work}). The core of the function is updating $L[i,j]$,
where $L[i,j]$ denotes the minimum sum of loads after partitioning the first
$i$ nodes into a total number of $j$ partitions. Let $C[i,k]$ denote the sum
of loads after partitioning the $i$th node into $k$ partitions. The value of
$C[i,k]$ can be obtained by calling a function similar to
\textsf{PartitionNode} except that the node is not really partitioned. Then we
can compute $L[i,j]$ by using the equation \[ L[i,j]\leftarrow min_{1\leq
k\leq j-i+1}\{L[i-1,j-k]+C[i,k]\}.\] Based on the arrays $L$ and $C$, the
  function can construct a map which maps from each node to the corresponding
  number of partitions.  

\paratitle{Node partitioning.} 
In Algorithm~\ref{alg:part_work}, we need to partition a node, if the number
of nodes is smaller than the number of workers (lines 13--16), or the load
balancing constraint cannot be satisfied (lines 23--27). Function
\textsf{PartitionNode} takes as input the node $n$ to be split, the
set of nodes $N_t$, the set of nodes $N_s$, and the number of splits $p$. It
outputs $p$ new nodes by partitioning node $n$.  If node $n$ belongs to $N_t$,
it indicates that the text similarity between objects and queries in $n$ is
small, and we partition $n$ into $p$ splits using text-partitioning. Otherwise,
we compare the workloads produced by using text-partitioning or
space-partitioning to partition $n$, and select the strategy resulting in a
smaller workload.

\paratitle{Node merging.} When the number of leaf nodes in the kd$^t$-tree
exceeds the number of workers $m$, we call function
\textsf{MergeNodesIntoPartitions} to divide those leaf nodes into $m$
partitions (line 18 of Algorithm~\ref{alg:part_work}). The function first
sorts the leaf nodes in descending order of their loads. In that order, for
each leaf node $n$, it finds the partition $part$ such that adding $n$ to
$part$ will result in a minimum load increase.  If assigning $n$ to $part$
does not increase the load balancing factor (i.e., $L_{max}/L_{min}$), $n$ is
assigned to $part$. Otherwise, $n$ is assigned to the partition that has
currently the smallest load.

\subsection{Index Structure on Dispatchers}
\label{sec:index}

The kd$^t$-tree can be used in the dispatcher to distribute the workload.  For
each spatio-textual object or updating request of an \STS queries, the
dispatcher obtains the corresponding leaf node(s) of the kd$^t$-tree by
traversing from the root node. The procedure takes $O(\log(m))$ time, which
may overload the dispatcher when the arrival speeds of objects and updating
requests of queries are very fast. 

\begin{figure}
\begin{minipage}[c]{\linewidth}
\center
\includegraphics[scale=0.5]{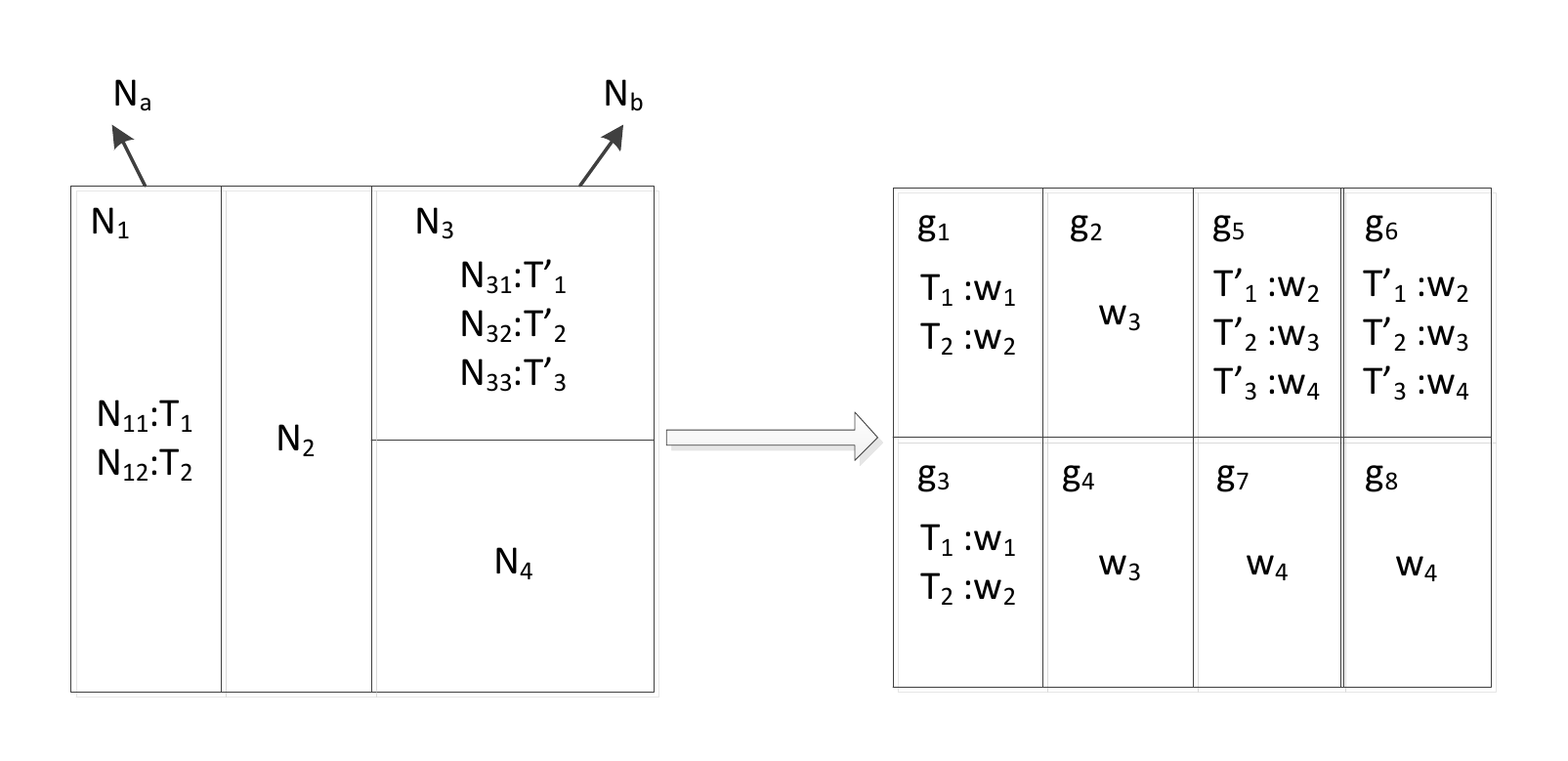}
\end{minipage}
\vspace{-2ex}
\caption{A grid$^t$ index transformed from a kd$^t$-tree}
\vspace{-3ex}
\label{fig:grid-disp}
\end{figure}

To alleviate the burden of the dispatcher, instead of maintaining a
kd$^t$-tree, we conduct the workload distribution using a grid$^t$ index with
each grid cell containing two hash maps $H_1$ and $H_2$. $H_1$ maps from terms
in the complete term set $T$ to the worker ids, and $H_2$ maps from terms in
\STS queries to the worker ids. To distribute a spatio-textual object, the
dispatcher first obtains the cell containing the object, and uses $H_2$ to
find the destination worker(s). The object can be discarded if it contains no
terms in $H_2$.  To distribute an updating request of an \STS query (note that
the request contains complete information of the \STS query), if the query
keywords are connected by AND operators only, the dispatcher obtains the cells
that query overlaps with, and looks up $H_1$ using the least frequent keyword
in each obtained cell to find the destination worker(s). $H_2$ is updated
correspondingly if it does not contain the keyword. For the query containing
OR operators, similar operations are conducted except that the set of the
least frequent keywords in each conjunctive norm form are used to find the
destination worker(s).
The grid$^t$ index can be built from the kd$^t$-tree.  The granularity of the
cell is decided by the leaf nodes of the kd$^t$-tree.
Figure~\ref{fig:grid-disp} shows an example of the relation between the
grid$^t$ index and the kd$^t$-tree in Figure~\ref{fig:kd-t}.

\subsection{Query Processing on Workers}
\label{sec:index_structure_on_workers}

Each worker maintains an in-memory index structure to organize the \STS
queries for accelerating the matching of incoming spatio-textual objects. The
index should be efficient in both matching \STS queries and spatio-textual
objects and updating of \STS queries. In our system, we adopt an index
structure named \textbf{G}rid-\textbf{I}nverted-\textbf{I}ndex (or GI$^2$ for
short)~\cite{st}. We choose GI$^2$ due to its efficiency in construction and
maintaining, which is important for processing a dynamic workload like the
data stream. Note that our system can be extended to adopt other index
structures.

After dividing the \STS queries by the cells they overlap with, GI$^2$
constructs an inverted index for each cell to organize the \STS queries. For
the query containing AND operators only, it is appended to the inverted list
of the least frequent keyword. For the query containing OR operators, it is
appended to the inverted lists of the least frequent keywords in each
conjunctive norm form.  The granularity of the cells can be determined
empirically in experiments. 

When processing a spatio-textual object $o$, the worker first looks up which
cell contains $o$ and then checks the associated inverted list in the cell for
each distinct term in $o.text$ to see whether $o$ can be matched by any \STS
query $q_i$ in that list. If there is a match, object $o$ is a result of
$q_i$.

To delete an \STS query from GI$^2$, we adopt the \textit{lazy-deletion}
strategy.  Specifically, we do not delete an \STS query immediately after
receiving the deletion request.  Instead, we record the ids of the queries to
be deleted into a hash table and remove the queries during the object
matching procedure.  In particular, while traversing an inverted list, we
check whether the id of each query in the inverted list appears in the hash
table. If yes, the query can be deleted.

\section{Dynamic Load Adjustment}
\label{sec:dynamic_load}

The dynamic property of the data stream gives rise to a changing workload.
The load of workers may change gradually over time. To handle
this, we conduct two types of load adjustments: local load adjustment and
global load adjustment. 

\subsection{Local Load Adjustment}
\label{local}

When the dispatcher detects that the load balancing constraint is violated, it
notifies the most loaded worker $w_o$ to transfer part of its workload to the
least loaded worker $w_l$. We expect the load adjustment procedure having the
following two features: 1.) The total amount of workload can be reduced after
the load adjustment. 2.) The migration cost is small so that the load
adjustment can be conducted efficiently. Existing
systems~\cite{gulisano2010streamcloud, mahmood2015tornado,
basik2015s, barazzutti2014elastic} do not support dynamic load adjustments that
consider both features. To achieve the goal, we propose new load adjustment
algorithms.

\paratitle{Algorithm overview.} We adjust the workload by migrating \STS
queries.  The queries are migrated in the unit of one cell in the grid$^t$
index.  Our load adjustment algorithm is composed of two phases. In the first
phase, we check whether some cells in $w_o$ and $w_l$ can be split
or merged so that the total amount of workload can be reduced. We conduct
related migration operations if such cells exist.  In the second phase, if the
load balancing constraint is still violated, we continue to select a set of
cells in $w_o$ to be migrated to $w_l$, with the goals of minimizing the
migration cost and the load balancing constraint can be satisfied.

\paratitle{Phase I.} We first check the $p$ most loaded cells $G_p$ in $w_o$,
where $p$ is a small parameter. For each cell $g_s$ not using
text-partitioning in $G_p$, we do the following checking: after using
text-partitioning to partition $g_s$ into two new cells $g_1$ and $g_2$,
whether the total amount of workload can be reduced if $g_1$ or $g_2$ is migrated to $w_l$.
If yes, we conduct the text-partitioning on $g_s$ and migrate the cell having
a smaller size between $g_1$ and $g_2$ to $w_l$. For each cell $g_t$ using
text-partitioning in $G_p$, if there exists a cell $g_{t}^{'}$ in $w_l$ which
shares the same space region as $g_t$ has, we check whether migrating $g_t$ to
$w_l$ and merging $g_t$ and $g_{t}^{'}$ can reduce the total load. If yes, we
conduct the migration.

\paratitle{Phase II.} If the load balancing constraint is still violated, we
conduct the second phase, which solves the Minimum Cost Migration problem.
Definition~\ref{def:grid_load} defines the load of a cell, and
Definition~\ref{def:min_cost_mig} presents the definition of the Minimum Cost
Migration problem.

\begin{definition}
  \label{def:grid_load}
  Given a time period, the load of a cell $g$ for the period is computed by
  \[L_g= n_o \cdot n_q,\] where $n_o$ is the number of
  spatio-textual objects falling in cell $g$ and $n_q$ is the average number of \STS
  queries stored in cell $g$ in the given time period.\endofproof
\end{definition}
\begin{definition}
\label{def:min_cost_mig}
\textbf{Minimum Cost Migration}:
Let $\tau$ denote the amount of load to be migrated. Consider a worker
$w_o$ with the set of cells $G_o$. The minimum cost migration problem is to
compute a set of cells $G_s$ to be migrated from worker $w_o$.
\[ G_s= \underset{G_s}{\arg\min} \ \ \sum_{g\in G_s}S_g \ \ \textrm{s.t. } \]
\[\sum_{g\in {G_s}}L_g \geq \tau, \]
where $S_g$ is the total size of
the queries in cell $g$.
\endofproof
\end{definition}
\begin{theorem}
  \label{th:min-cost-mig}
  The Minimum Cost Migration problem is NP-hard.
\end{theorem}
\noindent\textbf{Proof}: We prove by reducing from the Partition Problem,
which is NP-hard. In the Partition problem, given $n$ numbers $a_1, a_2,
\cdots, a_n\in \mathbf{N}$, we decide whether there exists a set $S\subseteq
\{1, 2, \cdots , n\}$ such that $\sum_{i\in S}a_i = \sum_{i\notin S}a_i$.  Let
MCMP($X$, $Y$) denote the decision version of the Minimum Cost Migration
problem, and the solution to it returns ``True'' if there exists such subset
of cells $G_s$ satisfying that $\sum_{g\in {G_s}}L_g
\geq X$ and $\sum_{g\in G_s}S_g=Y$, and returns ``False'' otherwise.  Given an
arbitrary instance of the Partition problem, we create an instance of
MCMP($X$, $Y$) as follows:

First, we produce the workload that for each cell $g_i$, $n_o=1$ and $n_q =
a_i$ (i.e., $L_{g_i}=a_i$), and $S_g=a_i$.

Second, we set $X=Y=\frac{\sum_{i=1}^n a_i}{2}$.

Therefore, the solution returns ``True'' iff there exists a set $S\subseteq \{1,
2, \cdots , n\}$ such that $\sum_{i\in S}a_i = \sum_{i\notin S}a_i$, and
returns ``False'' otherwise.  \endofproof

\subsubsection{Dynamic Programming Algorithm}
\label{sec:dp_alg}

We develop a dynamic programming algorithm to solve the Minimum Cost Migration
problem. Let $H(i,j)$ denote a subset of cells $\{g_1, g_2, \cdots , g_i\}$
such that the total size of its cells is no larger than $j$ and the total
load is maximum  \[ H(i,j) = \underset{H \subseteq\{g_1, g_2,\cdots
, g_i\} \land \sum_{g\in H}S_g \leq j }{\arg\max} \sum_{g\in H}L_g. \]

Let $A(i,j)$ denote the load of $H(i,j)$. We compute $A(i,j)$ for all $i\in
\{1, 2, \cdots , n\}$ and $j\in (0, P]$, where $P$ is an upper bound of the
minimum migration cost. Following shows the computation of $A(i,j)$.
\[
A(i,j)=\begin{cases}
  A(i-1,j) & \text{ if } j\leq S_{g_i}\\
max\{A(i-1,j), \\
  \qquad A(i-1,j-S_{g_i})+L_{g_i}\} & \text{ otherwise }
\end{cases}
.\]

The time complexity of the dynamic programming algorithm is $O(nP)$ and it
takes long running time when the size of cells is large. Moreover, it needs
$O(nP)$ memory space, which is expensive. To address these issues, we next
propose Algorithm \textsf{GR}, which is more efficient and requires much less
memory space.

\subsubsection{Greedy Algorithm}
\label{sec:greedy_alg}

We introduce the basic idea of the proposed greedy algorithm, denoted by
\textsf{GR}.  For each cell $g$, we compute $S_g/L_g$, which indicates the
relative cost of migrating the cell.  It is expected that migrating a cell
with a smaller relative cost may have smaller migration cost.  We scan cells
in $G_o$ in ascending order of their relative costs to find a set of cells to
be migrated.  For each cell we scan, if the total migrated load after
including it in the result is still less than $\tau$, the cell will be
included in the result, and we mark it by ``GS''; otherwise, the cell becomes a
candidate cell to be included in the result, and we mark it by ``GL".
The candidate cell will be included in the result, if in the subsequent scan  we
cannot find cells that can meet the load requirement and incur smaller migration
cost.

\begin{figure}[!ht]
\begin{minipage}[c]{\linewidth}
\includegraphics[width=\columnwidth]{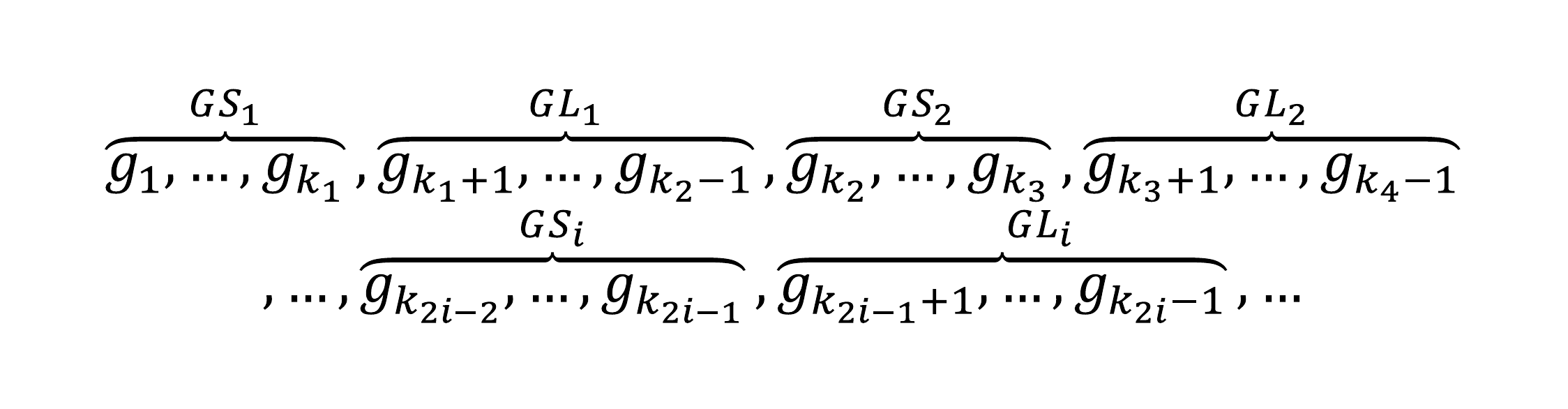}
\end{minipage}
\caption{Greedy algorithm for computing the set of cells to be migrated}
\vspace{-3ex}
\label{fig:lisifig1}
\end{figure}

We illustrate this process with Figure~\ref{fig:lisifig1}, where the cells are
sorted by their relative costs, and they are marked either by  ``GS'' or ``GL".
For any value of  $t$, we have 	$\sum_{i=1}^{t}\sum_{g\in GS_i}L_g < \tau$;
and $ \forall g^{'} \in GL_t, \sum_{i=1}^{t}\sum_{g\in GS_i}L_g + L_{g^{'}}
\geq \tau$.  For any  value of $t$, the set $GS_1 \cup GS_2 \cup \dots \cup
GS_t \cup \{g'\}$ is a candidate solution to the Minimum Cost Migration
problem.

During the scan, we  find the value for $t$ and the cell $g'$ such that $GS_1
\cup GS_2 \cup \dots \cup GS_t \cup \{g'\}$ has the minimum migration cost
among all the candidate solutions that we have scanned.

\subsection{Global Load Adjustment}
\label{sec:global}

The performance of current workload partitioning strategy will degrade when
there exists a significant change in the data distribution. To handle this, we
periodically check whether a workload repartitioning is necessary on a recent
sample of data. If yes, we conduct a workload repartitioning using the
algorithm presented in Section~\ref{sec:workload_part}. To avoid the large
migration cost after the workload repartitioning, we make a temporary
compromise on the system performance by maintaining two workload distribution
strategies: one for the old \STS queries before the workload repartitioning,
and the other one for the new \STS queries. When the amount of old \STS
queries becomes small, we conduct the migration and stop the old workload
distribution strategy.  We set a long period for the checking, e.g., once per
day, which is reasonable as the data distribution usually changes slowly.

\section{Experimental Evaluation}
\label{sec:experimental_evaluation}

\subsection{Experimental Setup}
\label{sec:experimental_setup}

We build \PSS on top of Storm\footnote{http://storm.apache.org}, an open-source stream
processing engine.  Our system is deployed on the Amazon EC2 platform using a
cluster of 32 c3.large instances. Each c3.large instance has 2 vCPUs running
Intel Xeon E5-2680 at 2.5GHz and 3.75 GB RAM. The network performance is
moderate and no further information about the bandwidth is provided by Amazon.
We rent another high performance m4.2xlarge instance (which is equipped
with 8 vCPUs and 32 GB RAM, and has high IO performance) to emit streams of
spatio-textual objects and \STS queries to our system. 

\smallskip\noindent\textbf{Datasets and \STS Queries.} Our experiments are
conducted on two datasets, which are TWEETS-US and TWEETS-UK.  TWEETS-US
consists of 280 million spatio-textual tweets in America and TWEETS-UK
consists of 58 million spatio-textual tweets in Britain.  Since we do not have
real-life \STS queries, we synthesize queries based on tweets. 
The number of keywords is a random number ranging from 1 to 3 and the keywords
are connected by either AND or OR operators. The query range is a square and
the center is randomly selected from the locations of tweets in TWEETS-US or
TWEETS-UK. We synthesize four groups of \STS queries as follows.

(1) STS-US-Q1, STS-US-Q2: The two groups of queries are synthesized based on
TWEETS-US. For STS-US-Q1, the side lengths of the rectangle are randomly
assigned between 1km and 50km and the keywords share the same distribution as
the terms in TWEETS-US, i.e., the keywords in queries satisfy the power-law
distribution. For STS-US-Q2, the side lengths of the rectangle are
randomly assigned between 1km and 100km and the queries contain at least one
keyword that is not in the top 1\% most frequent terms in TWEETS-US.
      
(2) STS-UK-Q1, STS-UK-Q2: The two groups of queries are synthesized
based on TWEETS-UK as we do for TWEETS-US.

\smallskip\noindent\textbf{Workload.} The ratio of processing a spatio-textual
tweet to inserting or deleting an \STS query is approximately 5. The arrival
speeds of requests for inserting an \STS query and deleting an \STS query are
equivalent.  It indicates that the number of \STS queries in the system will
be stable eventually.  We use a parameter $\mu$ to control the number of \STS
queries. We achieve this by using a Gaussian distribution $\mathcal{N}(\mu,
\sigma ^2)$ to determine the number of newly arrived \STS queries between
inserting an STS query and deleting it. We set different values of $\mu$ in
the experiments with $\sigma=0.2 \mu$.

\subsection{Evaluating Baselines}
\label{sec:experimental_results_framework}

We evaluate the performance of baseline workload distribution algorithms. 

\noindent\textbf{Text-Partitioning}. \textit{Text-partitioning} algorithms
divide the lexicon into several partitions, each of which is assigned to one
worker, and distribute the workload based on the textual content of
objects/queries. We implement the following three \textit{text-partitioning}
algorithms: (1) Algorithm \textit{frequency-based partitioning} uses the
frequency values of the terms to do the partitioning.  (2) Algorithm
\textit{hypergraph-based partitioning}~\cite{cambazoglu2013term} constructs a
hypergraph based on the co-occurrence of terms and partitions that hypergraph.
(3) Algorithm \textit{metric-based partitioning}~\cite{basik2015s} uses a
metric function to do the partitioning.

\noindent\textbf{Space-Partitioning}. \textit{Space-partitioning} algorithms
divide the space into several partitions, each of which is assigned to one
worker, and distribute the workload based on the spatial content of
objects/queries.  We implement the following three \textit{space-partitioning}
algorithms: (1) Algorithm \textit{grid
partitioning}~\cite{eldawy2015spatialhadoop} represents the space as a set
of cells and then partitions that set of cells. We set its granularity as
$2^6\times 2^6$, as it performs best under that granularity. (2) Algorithm
\textit{kd-tree partitioning}~\cite{aly2015aqwa,mahmood2015tornado} constructs
a kd-tree to do the partitioning, of which each leaf node represents one
partition. We transform the kd-tree to a grid index to accelerate the workload
distribution in the dispatchers. (3) Algorithm \textit{R-tree
partitioning}~\cite{eldawy2015spatialhadoop} constructs a R-tree to do the
partitioning, and then partitions the set of leaf nodes.

Note that this is the first work on empirically comparing the performance of
these algorithms for handling a spatio-textual data stream.  These experiments are
conducted using 4 dispatchers and 8 workers. The
granularity of GI$^2$ index of workers in all algorithms is set as $2^6\times
2^6$ for fairness.

\begin{figure}
\begin{minipage}[c]{\linewidth}
\subfigure[\scriptsize Text-Partitioning (\#Q1=5M)]
{\label{fig:basic_th_text_Q1}\includegraphics[scale=0.33]{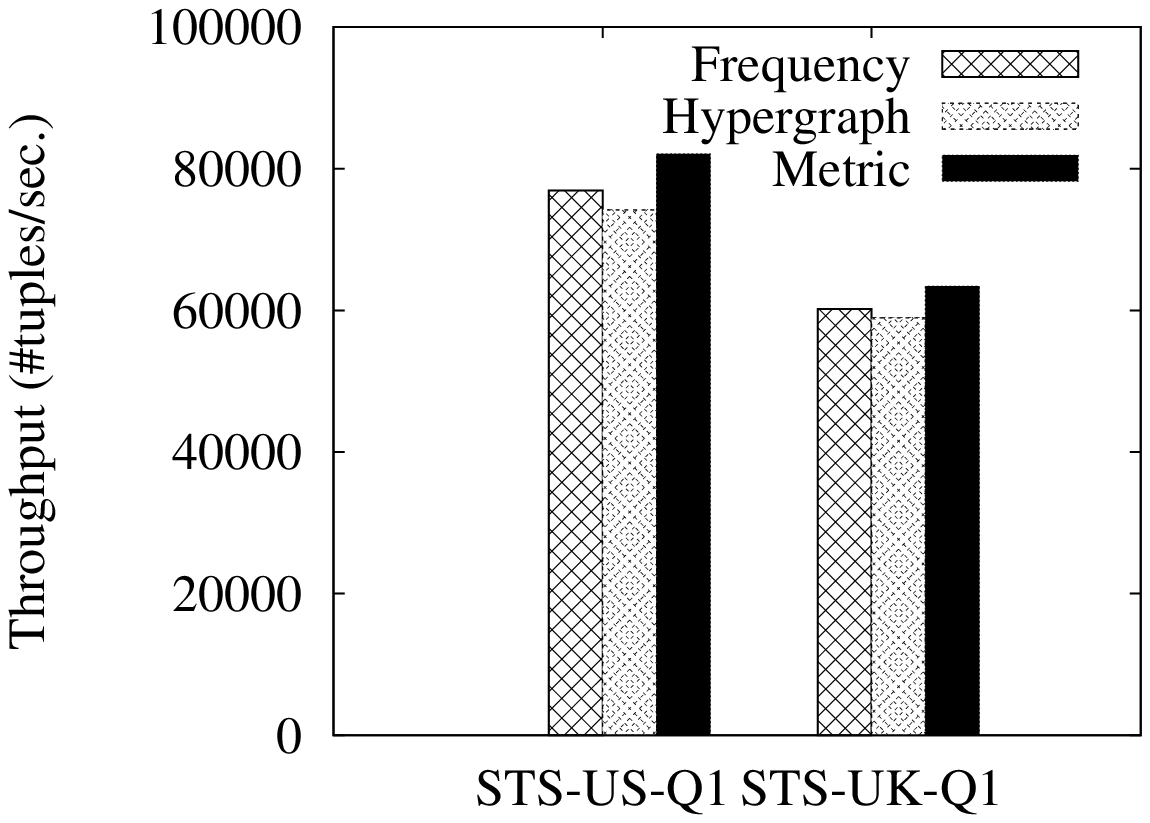}}
\subfigure[\scriptsize Text-Partitioning (\#Q2=10M)]
{\label{fig:basic_th_text_Q2}\includegraphics[scale=0.33]{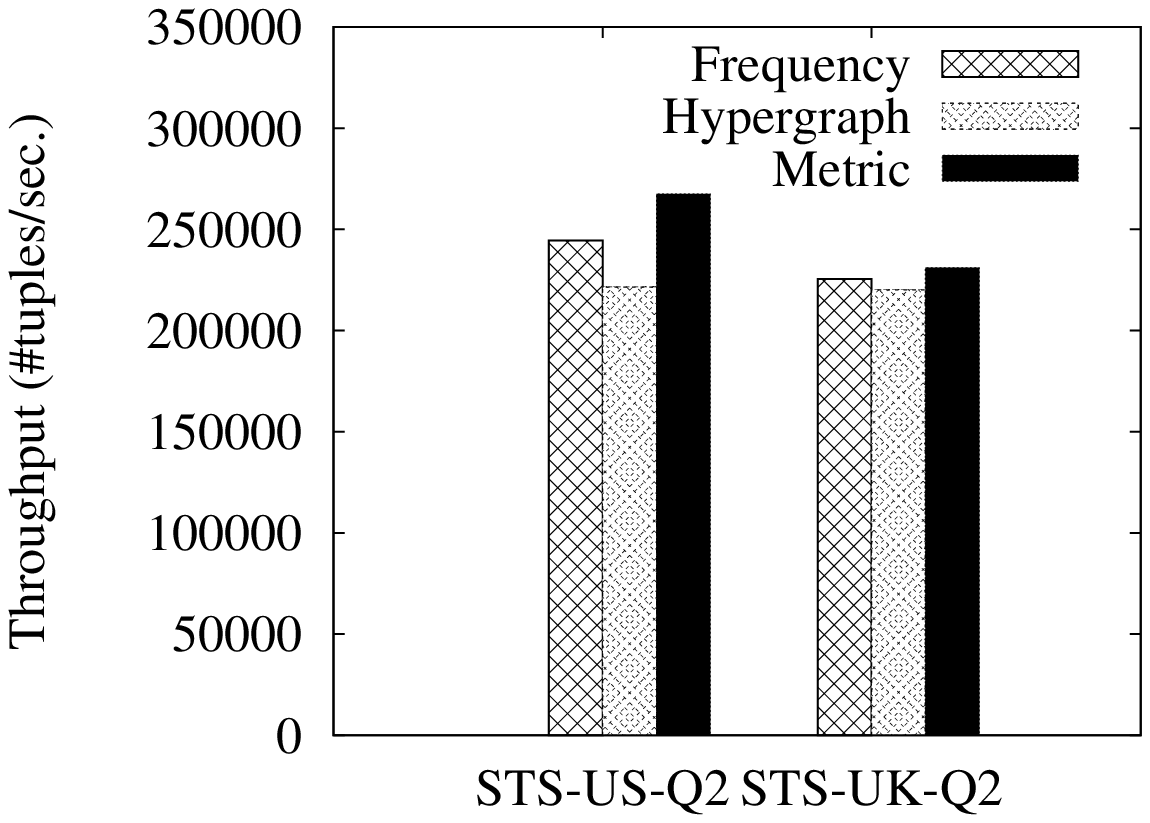}}
\subfigure[\scriptsize Space-Partitioning (\#Q1=5M)]
{\label{fig:basic_th_space_Q1}\includegraphics[scale=0.33]{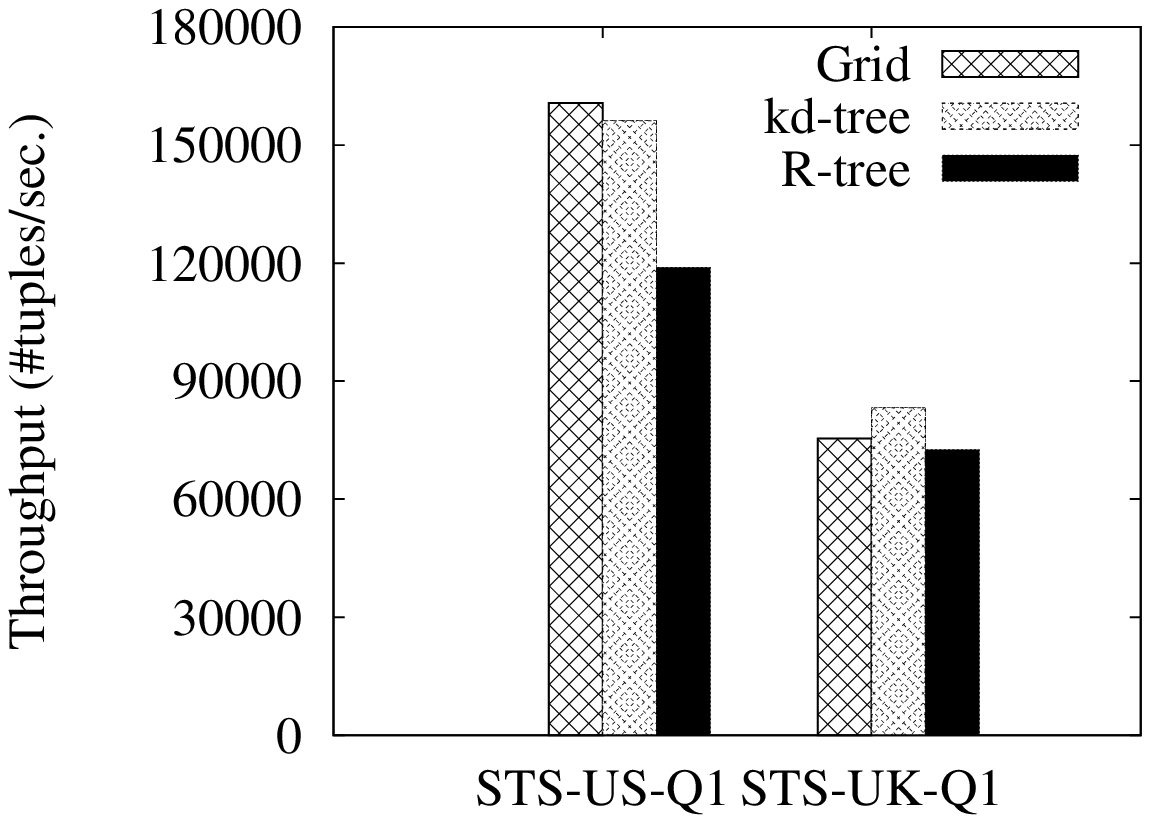}}
\hspace{0.2cm}
\subfigure[\scriptsize Space-Partitioning (\#Q2=10M)]
{\label{fig:basic_th_space_Q2}\includegraphics[scale=0.33]{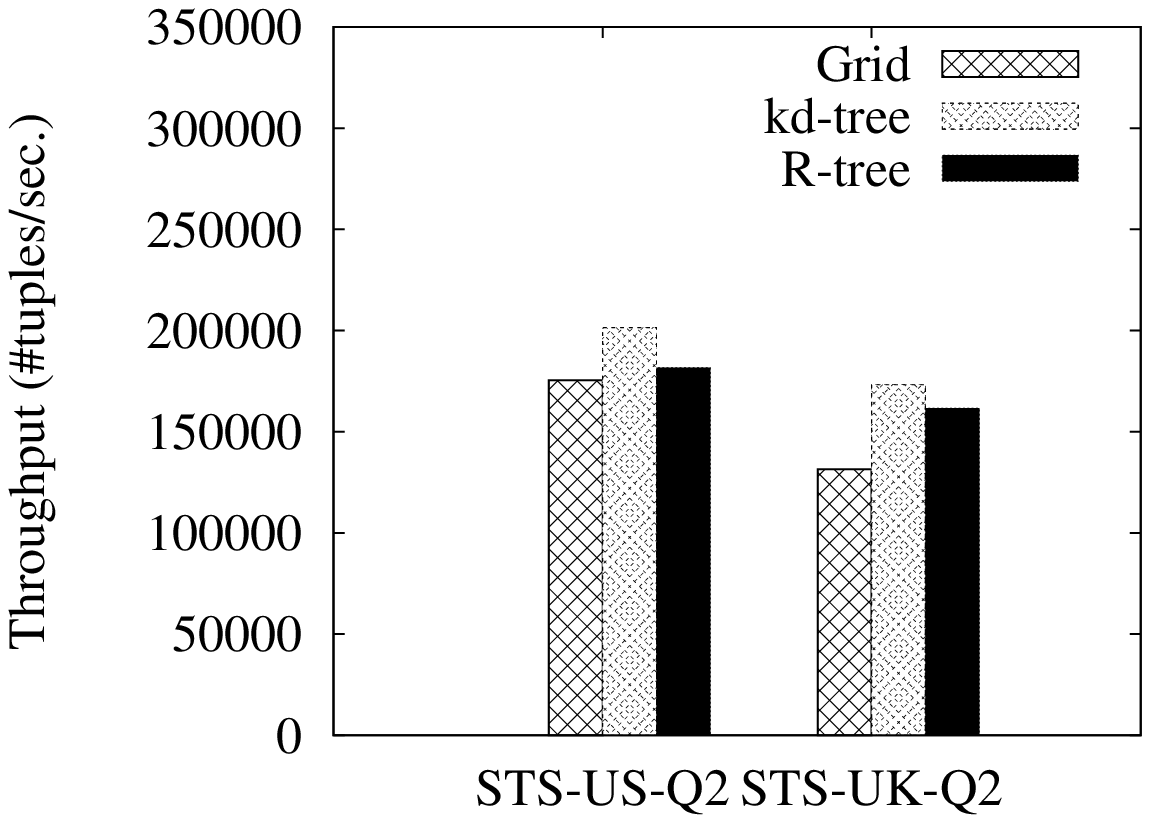}}
\vspace{-1ex}
\caption{Throughput Comparison (Baselines)}
\label{fig:throughput_basic}
\vspace{-5ex}
\end{minipage}
\end{figure}

\paratitle{Throughput} The throughput is the average number of tuples (i.e.,
including both query insertion/deletion and object matching) being processed
per second when the processing capacity of the system is reached (i.e., the
input speed of the data stream is tuned to be closed to the maximum number of
tuples that the system can handle in unit time).  We set $\mu=5$M for Q1
queries, and set $\mu=10$M for Q2 queries.

Figure~\ref{fig:throughput_basic} shows the throughputs of the baselines. For
Q1 queries, \textit{space-partitioning} algorithms perform better
than \textit{text-partitioning} algorithms. For example, for STS-US-Q1, the
throughput of \textit{grid partitioning} is 160,659 tuples/second, and the
throughput of \textit{frequency-based partitioning} is 76,921 tuples/second.
The reason is that \textit{space-partitioning} algorithms impose a smaller
amount of workload on the workers: the \textit{text-partitioning} algorithms
send each object to multiple workers as the keywords in Q1 queries are
frequent in objects. For Q2 queries, \textit{text-partitioning} algorithms
perform better than \textit{space-partitioning} algorithms. For
example, for STS-US-Q2, the throughput of \textit{metric-based partitioning}
is 267,415 tuples/second, and the throughput of \textit{R-tree partitioning}
is 181,542 tuples/second.  The reason is that the Q2 queries have larger query
ranges and less frequent keywords, and therefore \textit{text-partitioning}
algorithms impose a smaller workload on the workers. 

Overall, \textit{metric-based partitioning} performs the best among
\textit{text-partitioning} algorithms, and \textit{kd-tree
partitioning} performs the best among \textit{space-partitioning} algorithms. 
Therefore, we select \textit{metric-based partitioning} and \textit{kd-tree
partitioning} for further evaluation.

\subsection{Evaluating Hybrid Partitioning}
\label{sec:evaluate_hybrid}

We evaluate the performance of our hybrid partitioning algorithm in this set
of experiments. To simulate the situation that users in different regions have
different preferences over the spatio-textual objects, we synthesize two new
groups of \STS queries: STS-US-Q3 and STS-UK-Q3. STS-US-Q3 is synthesized by
partitioning the spatial range of the USA into 100 regions of equal size, and
for each region we use STS-US-Q1 or STS-US-Q2. Similar procedure is conducted
to synthesize STS-UK-Q3

We first evaluate \textit{hybrid partitioning} using 4 dispatchers and 8
workers. Then we evaluate the scalability of \textit{hybrid partitioning}
using 4 dispatchers while varying the number of workers.

\begin{figure}
\begin{minipage}[c]{\linewidth}
  \centering
\subfigure[\scriptsize \#Queries=5M (Q1)]
{\label{fig:hybrid_th_q1}\includegraphics[scale=0.4]{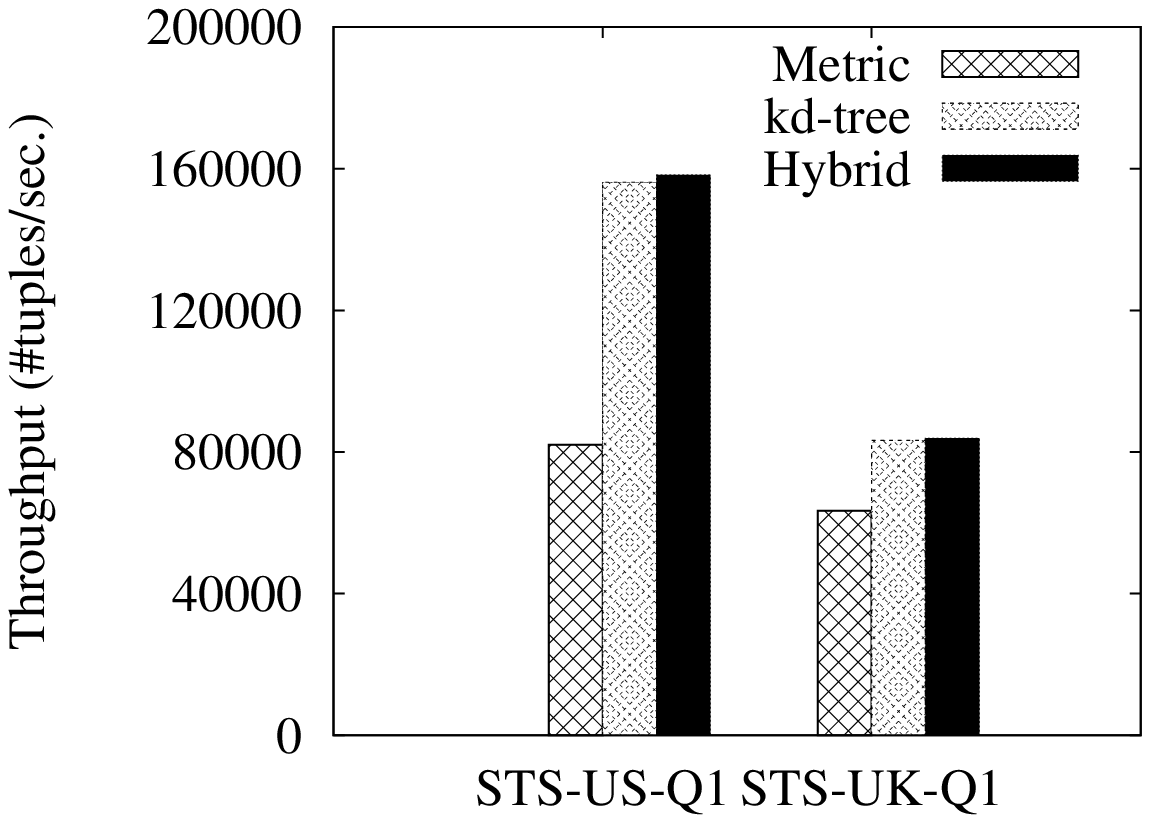}}
\subfigure[\scriptsize \#Queries=10M (Q2)]
{\label{fig:hybrid_th_q2}\includegraphics[scale=0.4]{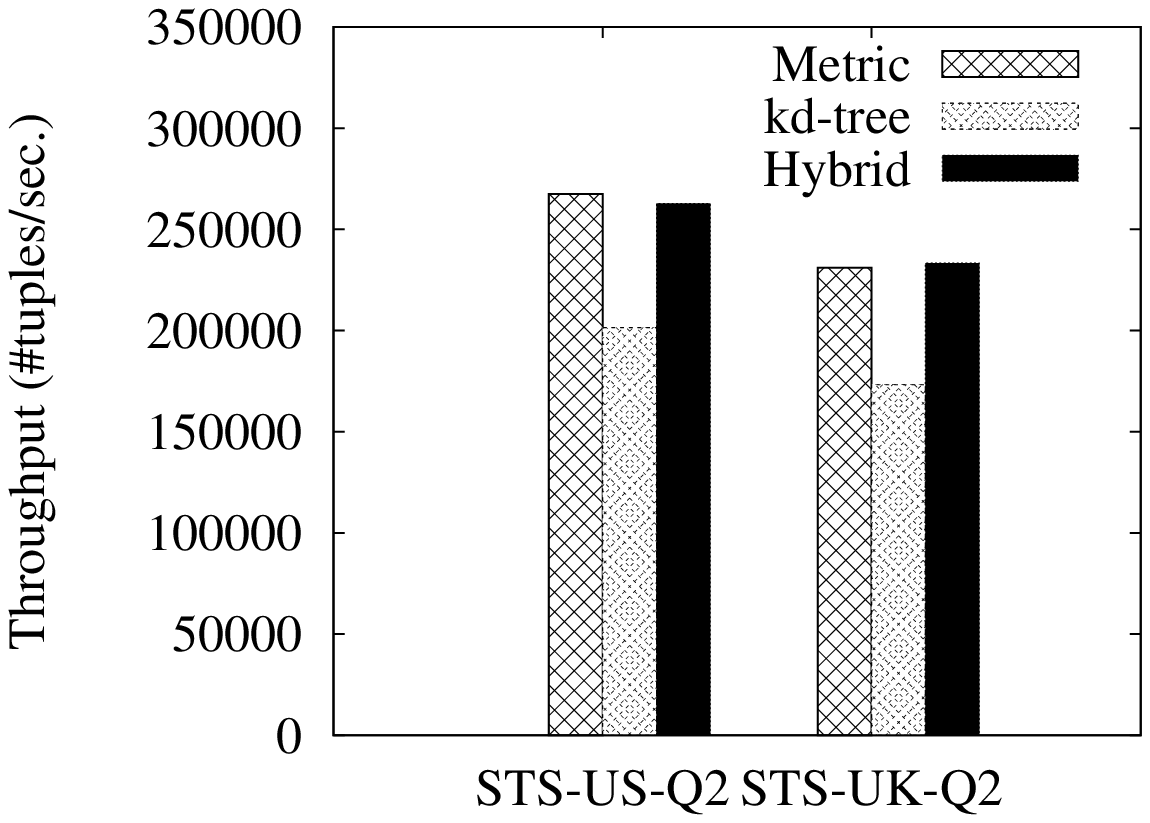}}
\subfigure[\scriptsize \#Queries=10M (Q3)]
{\label{fig:hybrid_th_q1_q2}\includegraphics[scale=0.4]{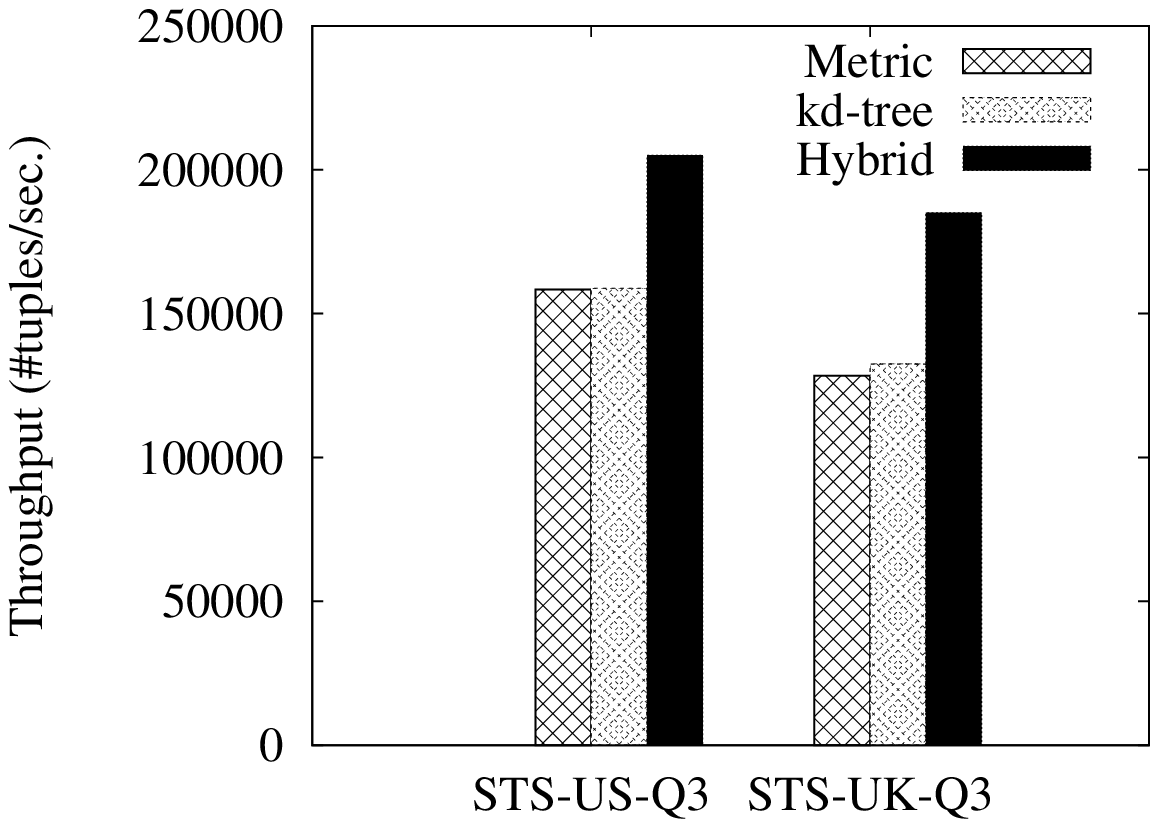}}
\vspace{-1ex}
\caption{Throughput Comparison}
\vspace{-3ex}
\label{fig:throughput_hybrid}
\end{minipage}
\end{figure}

\paratitle{Throughput} Figure~\ref{fig:throughput_hybrid} shows the
throughputs of \textit{hybrid partitioning}, \textit{metric-based
partitioning} and \textit{kd-tree partitioning}. Algorithm \textit{hybrid
partitioning} shows the overall best performance regardless of the data
distributions. In Figure~\ref{fig:hybrid_th_q1_q2}, \textit{hybrid
partitioning} outperforms \textit{metric-based partitioning} and
\textit{kd-tree partitioning} by 30\% in terms of throughput.  In
Figure~\ref{fig:hybrid_th_q1}, for STS-US-Q1, \textit{hybrid partitioning}
slightly outperforms \textit{kd-tree partitioning}.  Both \textit{hybrid
partitioning} and \textit{kd-tree partitioning} perform much better than
\textit{metric-based partitioning}. It is because that most keywords in
STS-US-Q1 are frequent among objects, which results in a larger workload using
\textit{text-partitioning}. This can also explain why \textit{hybrid
partitioning} and \textit{kd-tree partitioning} have similar performance.
Because \textit{hybrid partitioning} wisely choose \textit{space-partitioning}
to partition the workload in most regions.  In Figure~\ref{fig:hybrid_th_q2},
\textit{hybrid partitioning} and \textit{metric-based partitioning} have a
larger throughput than \textit{kd-tree partitioning} does. This can be
explained in two aspects.  First, the larger query range results in a larger
workload for \textit{kd-tree partitioning}. Second, the keywords in STS-US-Q2
and STS-UK-Q2 are less frequent, which improves the performance of
\textit{text-partitioning}.  

\paratitle{Latency} The latency is the average time of each tuple staying in
the system. We evaluate all the algorithms using a moderate input speed of the
data stream.

\begin{figure}
\begin{minipage}[c]{\linewidth}
  \centering
\subfigure[\scriptsize \#Queries=5M (Q1)]
{\label{fig:hybrid_latency_q1}\includegraphics[scale=0.4]{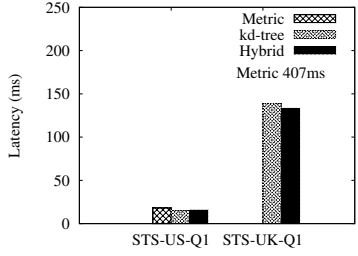}}
\subfigure[\scriptsize \#Queries=10M (Q2)]
{\label{fig:hybrid_latency_q2}\includegraphics[scale=0.4]{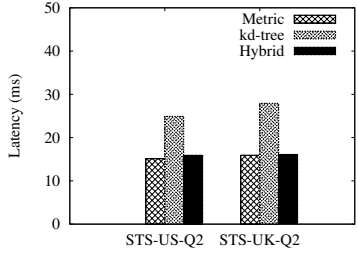}}
\subfigure[\scriptsize \#Queries=10M (Q3)]
{\label{fig:hybrid_latency_q1_q2}\includegraphics[scale=0.4]{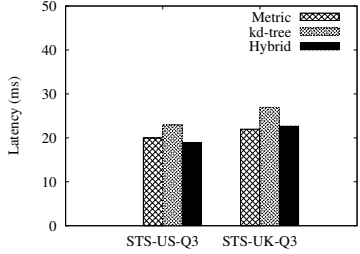}}
\vspace{-1ex}
\caption{Latency Comparison}
\vspace{-3ex}
\label{fig:latency_hybrid}
\end{minipage}
\end{figure}

Figure~\ref{fig:latency_hybrid} shows the latencies of \textit{hybrid
partitioning}, \textit{metric-based partitioning} and \textit{kd-tree
partitioning}. Algorithm \textit{hybrid partitioning} has smaller latency.
The latency of \textit{kd-tree partitioning} is noticeably larger than
\textit{hybrid partitioning} and \textit{metric-based partitioning} in
Figure~\ref{fig:hybrid_latency_q2}, e.g., 25ms versus 15ms.  This is caused by
the larger spatial ranges of queries in STS-US-Q2 and STS-UK-Q2.  Algorithm
\textit{metric-based partitioning} has similar latency values as
\textit{hybrid partitioning} does with the exception on STS-UK-Q1, which takes
407ms. This is caused by the poor performance of \textit{text-partitioning}
when query keywords are frequent.  Figure~\ref{fig:latency_hybrid}
demonstrates that \textit{hybrid partitioning} achieves small latency for
workload processing.

\begin{figure}
\begin{minipage}[c]{\linewidth}
  \centering
\subfigure[\scriptsize \#Queries=5M (Q1)]
{\label{fig:hybrid_mem_disp_q1}\includegraphics[scale=0.4]{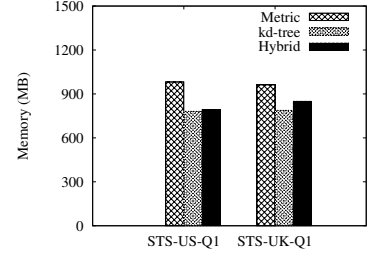}}
\subfigure[\scriptsize \#Queries=10M (Q2)]
{\label{fig:hybrid_mem_disp_q2}\includegraphics[scale=0.4]{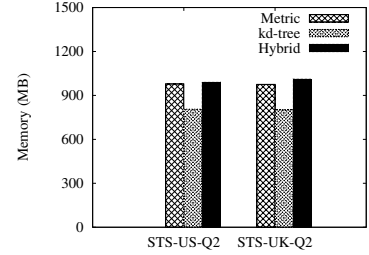}}
\subfigure[\scriptsize \#Queries=10M (Q3)]
{\label{fig:hybrid_mem_disp_q1_q2}\includegraphics[scale=0.4]{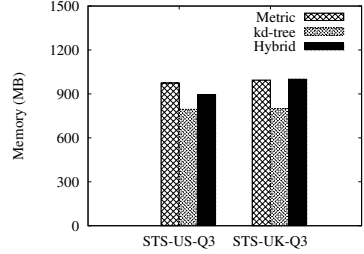}}
\vspace{-1ex}
\caption{Memory Comparison (Dispatchers)}
\vspace{-3ex}
\label{fig:mem_disp_hybrid}
\end{minipage}
\end{figure}

\paratitle{Memory} Figure~\ref{fig:mem_disp_hybrid} shows the average memory
usages of dispatchers of \textit{hybrid partitioning}, \textit{metric-based
partitioning} and \textit{kd-tree partitioning}.  Overall, the memory usage of
all the methods is not large, e.g., less than 1000MB, which is acceptable in
practice.  Algorithm \textit{kd-tree partitioning} uses less memory than
\textit{metric-based partitioning} and \textit{hybrid partitioning}. The
memory usage of \textit{hybrid partitioning} is the highest for Q2 queries
than for Q1 and Q3. This is because that, for Q2 queries, more cells in the
grid$^t$ index need to maintain additional text partitioning information.

\begin{figure}
\begin{minipage}[c]{\linewidth}
  \centering
\subfigure[\scriptsize \#Queries=5M (Q1)]
{\label{fig:hybrid_mem_worker_q1}\includegraphics[scale=0.4]{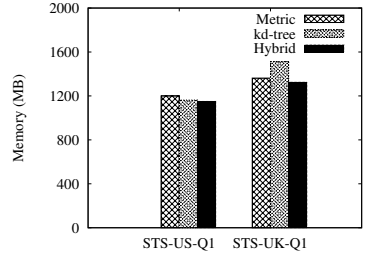}}
\subfigure[\scriptsize \#Queries=10M (Q2)]
{\label{fig:hybrid_mem_worker_q2}\includegraphics[scale=0.4]{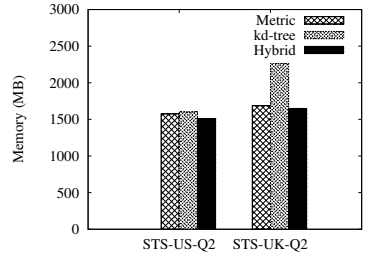}}
\subfigure[\scriptsize \#Queries=10M (Q3)]
{\label{fig:hybrid_mem_worker_q1_q2}\includegraphics[scale=0.4]{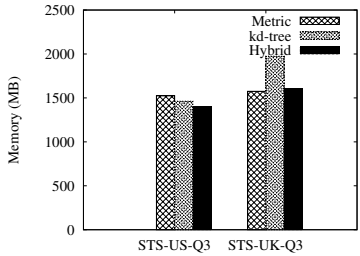}}
\vspace{-1ex}
\caption{Memory Comparison (Workers)}
\vspace{-3ex}
\label{fig:mem_worker_hybrid}
\end{minipage}
\end{figure}

Figure~\ref{fig:mem_worker_hybrid} shows the average memory usages of workers
of the three methods.  Algorithm \textit{hybrid partitioning} has an overall
best performance, which invokes the smallest memory usage in most cases. This
is due to the reason that \textit{hybrid partitioning} distributes \STS
queries to workers considering the data distributions in different regions,
which reduces the cases of one \STS query being stored in multiple workers. We
also observe that all the three methods do not impose large memory requirements
on workers.

\begin{figure}
\begin{minipage}[c]{\linewidth}
  \centering
\subfigure[\scriptsize \#Queries=10M (STS-UK-Q1)]
{\label{fig:scal_th_q1}\includegraphics[scale=0.4,angle=270]{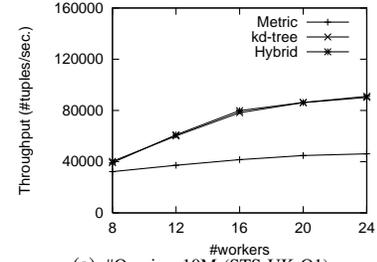}}
\subfigure[\scriptsize \#Queries=20M (STS-UK-Q2)]
{\label{fig:scal_th_q2}\includegraphics[scale=0.4,angle=270]{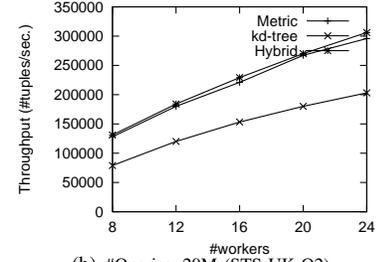}}
\subfigure[\scriptsize \#Queries=20M (STS-UK-Q3)]
{\label{fig:scal_th_q3}\includegraphics[scale=0.4,angle=270]{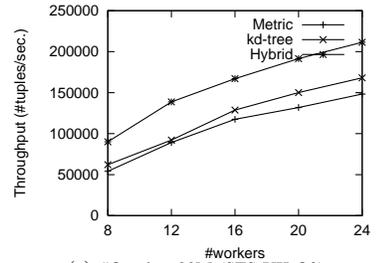}}
\vspace{-1ex}
\caption{Scalability}
\label{fig:scalability}
\end{minipage}
\end{figure}

\paratitle{Scalability} Figure~\ref{fig:scalability} shows the throughputs of
the evaluated algorithms as we vary the number of workers. As we can see,
\textit{hybrid partitioning} exhibits the best performance in most cases, and
scales well with the number of workers. Algorithm \textit{metric-based
partitioning} has the worst scalability in Figure~\ref{fig:scal_th_q1} as the
keywords in STS-UK-Q1 are frequent, which results in a larger workload on the
workers. Figure~\ref{fig:scal_th_q2} shows that \textit{kd-tree partitioning}
has the worst scalability.  

\subsection{Evaluating Dynamic Load Adjustment}
\label{sec:experimental_results_migration} 

\begin{figure}
\vspace{-2ex}
  \begin{minipage}[c]{\linewidth}
    \centering
      \subfigure[\scriptsize Time of Selecting Cells]
      {\label{fig:time_sele_cell_1m_us10}\includegraphics[scale=0.5, angle=270]{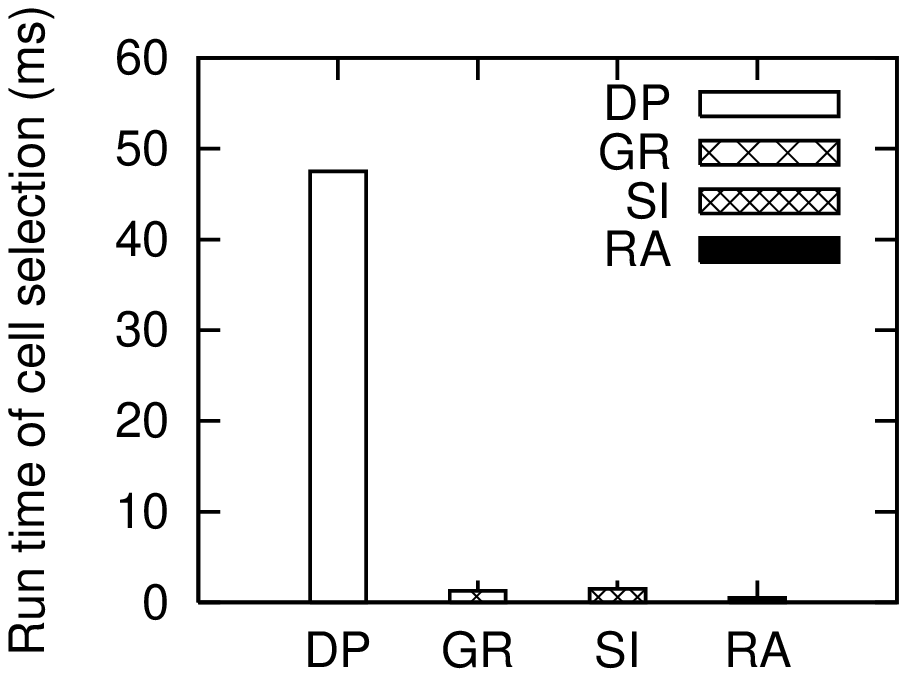}} 
      \label{fig:time_cell_selection_sts10}
      \subfigure[\scriptsize Migration Cost and Time]
      {\label{fig:mig_cost_time_1m_us10}\includegraphics[scale=0.50,
      angle=270]{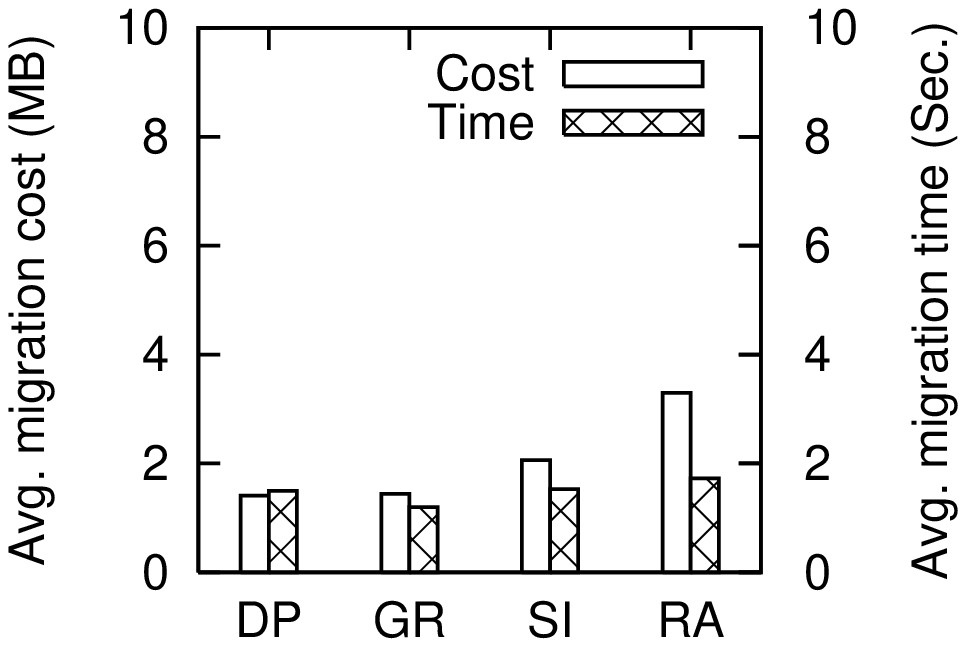}}
      \label{fig:migration_cost_time_sts10}
      \subfigure[\scriptsize Percentage of Latency Values]
      {\label{fig:mig_percent_1m_us10}\includegraphics[scale=0.50,
      angle=270]{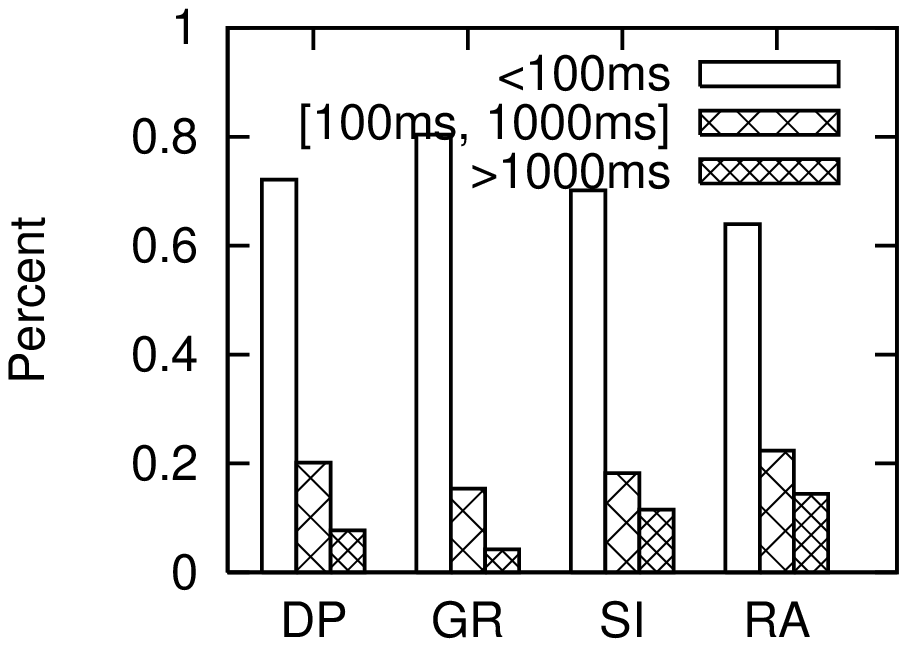}}
      \vspace{-1ex}
    \caption{Migration Experiments (\#Q=1M, STS-US-Q1)}
    \label{fig:migration_exps_sts10}
      \vspace{-3.5ex}
\end{minipage}
\end{figure}

In this set of experiments, we evaluate the performance of our dynamic load
adjustment algorithms. The experiments are conducted on 4 dispatchers and 8
workers. In addition to our proposed dynamic programming (\textbf{DP})
algorithm and greedy (\textbf{GR}) algorithm,
we implement two other algorithms for comparison. 

\textbf{SI}: It is another greedy algorithm which adds cells into $G_s$ (the
set of cells to be migrated), in descending order of their sizes. 

\textbf{RA}: It selects the cells to be migrated randomly.

\begin{figure}
\begin{minipage}[c]{\columnwidth}
\vspace{-1ex}
\subfigure[\scriptsize \#Queries=5M]
{\label{fig:time_sele_cell_5m_us50}\includegraphics[scale=0.42,
angle=270]{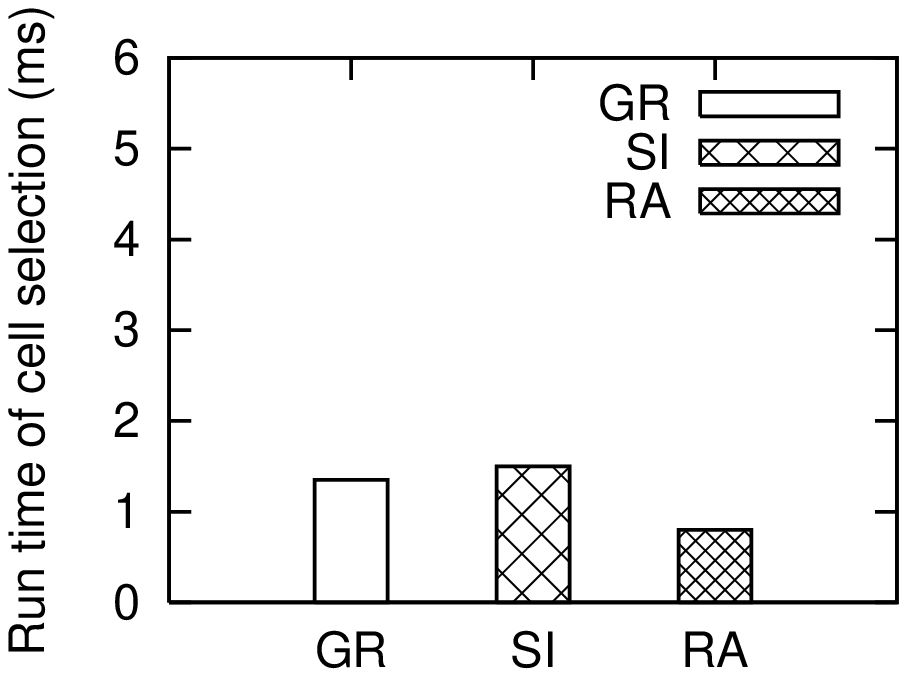}}
\subfigure[\scriptsize \#Queries=10M]
{\label{fig:time_sele_cell_10m_us50}\includegraphics[scale=0.42,
angle=270]{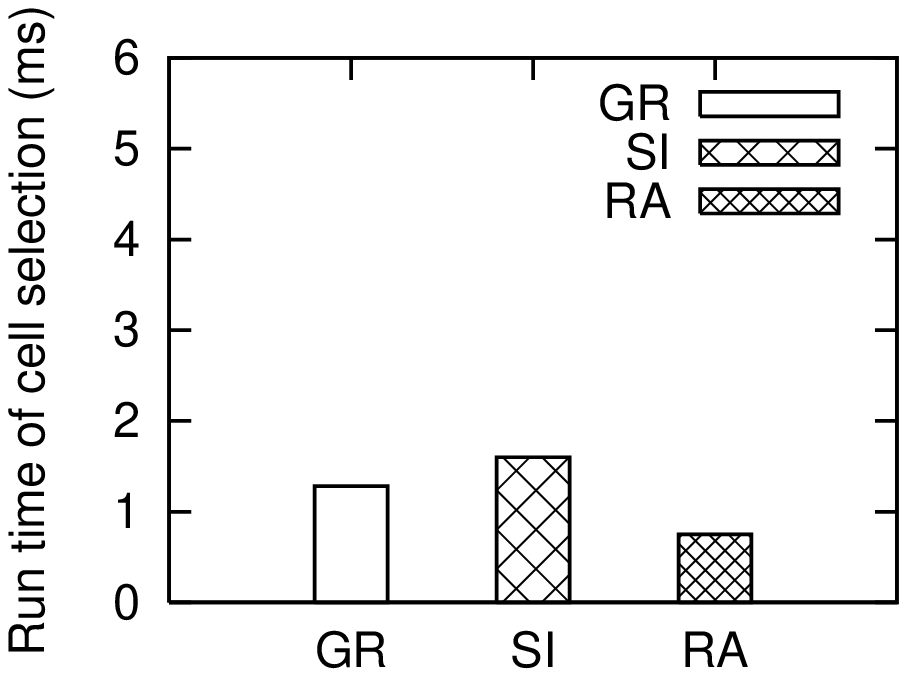}}
\vspace{-1ex}
\caption{Average Time of Selecting Cells (STS-US-Q1)}
\label{fig:time_cell_selection}
\end{minipage}
\vspace{-4ex}
\end{figure}

\paratitle{Comparing the running time of selecting cells for migration} We run
the four algorithms on the same worker (having the same set of cells) and
measure their running time for selecting cells for migration, and the results
are shown in Figure~\ref{fig:time_sele_cell_1m_us10} and
Figure~\ref{fig:time_cell_selection}. Note that workers run out of memory when
running DP on queries used for the experiments in
Figure~\ref{fig:time_cell_selection}. 
As shown in Figure~\ref{fig:time_sele_cell_1m_us10}, DP runs significantly
longer than the other algorithms. This is due to the high time complexity of
DP.  As expected, RA is the fastest among all the algorithms. However, both GR
and SI are very efficient too. By comparing
Figures~\ref{fig:time_sele_cell_5m_us50} and~\ref{fig:time_sele_cell_10m_us50}
, we can see that the running time of GR, SI and RA
does not change with the number of queries. The reason is that their
running time is only determined by the number of cells.

\paratitle{Comparing the migration cost} We conduct the evaluation by running
a workload consisting of processing a sample of spatio-textual tweets in 60
days and the insertions/deletions of \STS queries.
Since our tweets are only a small sample of real-life tweets, we scale out the
workload by reading 4 hours of tweets in every 10 seconds. Note that we
utilize the timestamps of tweets for the scale-out.  In each workload
migration, we measure the size of migrated data and the time in doing
migration. 

\begin{figure}
\vspace{-1ex}
\begin{minipage}[c]{\linewidth}
\subfigure[\scriptsize \#Queries=5M]
{\label{fig:mig_cost_time_5m_us50}\includegraphics[scale=0.42,
angle=270]{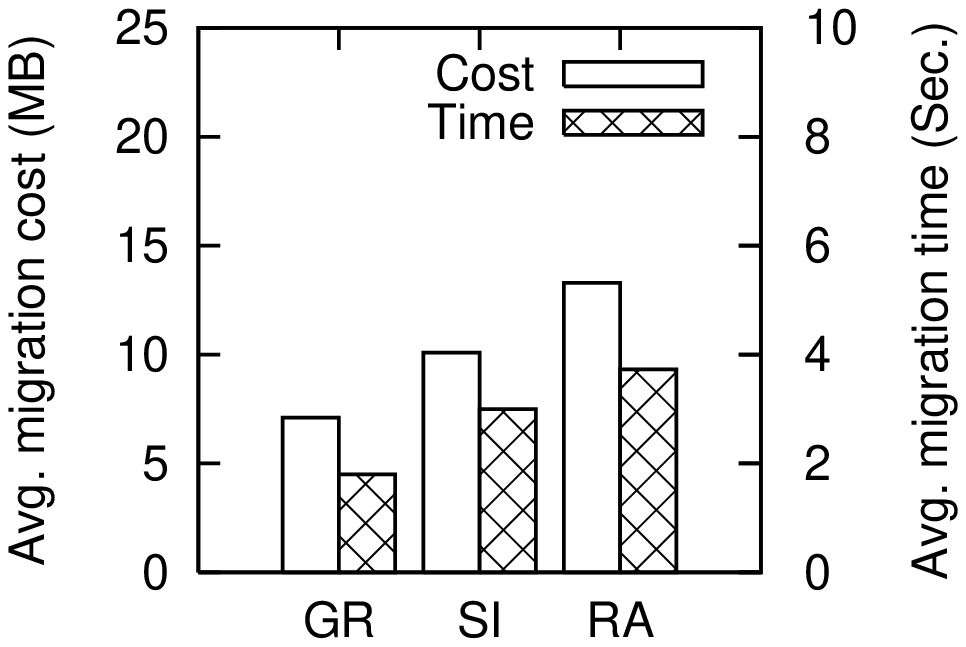}}
\subfigure[\scriptsize \#Queries=10M]
{\label{fig:mig_cost_time_10m_us50}\includegraphics[scale=0.42,
angle=270]{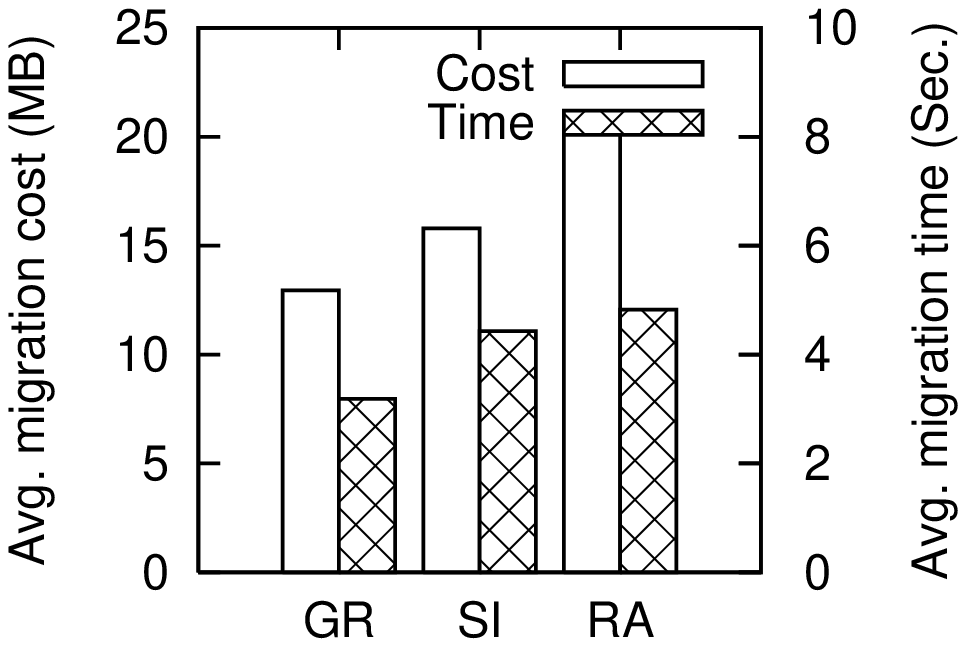}}
\vspace{-1ex}
\caption{Migration Cost and Time (STS-US-Q1)}
\label{fig:migration_cost_time}
\vspace{-1ex}
\end{minipage}

\begin{minipage}[c]{\linewidth}
\subfigure[\scriptsize \#Queries=5M]
{\label{fig:mig_percent_5m_us50}\includegraphics[scale=0.42,
angle=270]{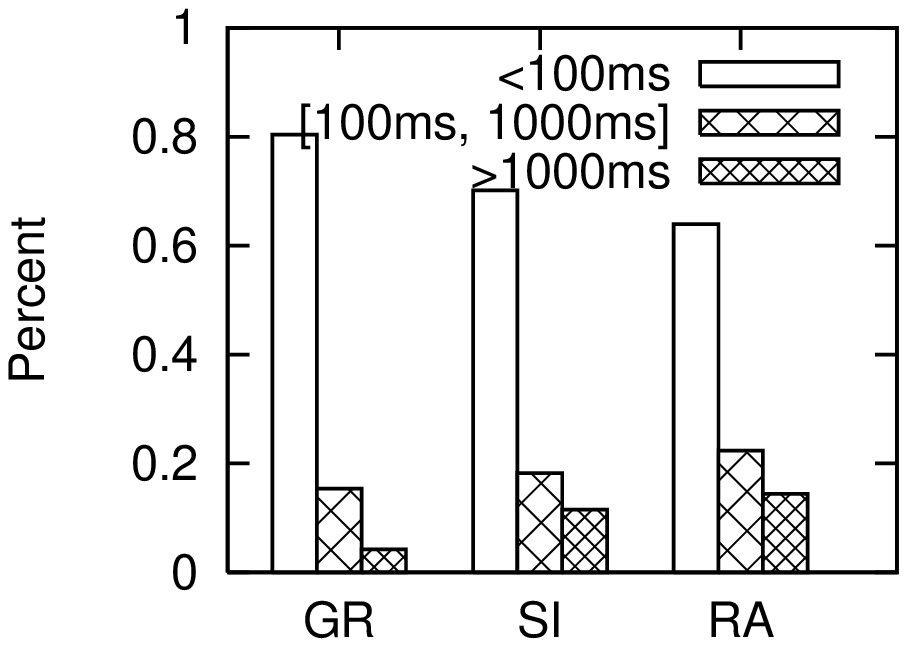}}
\subfigure[\scriptsize \#Queries=10M]
{\label{fig:mig_percent_10m_us50}\includegraphics[scale=0.42,
angle=270]{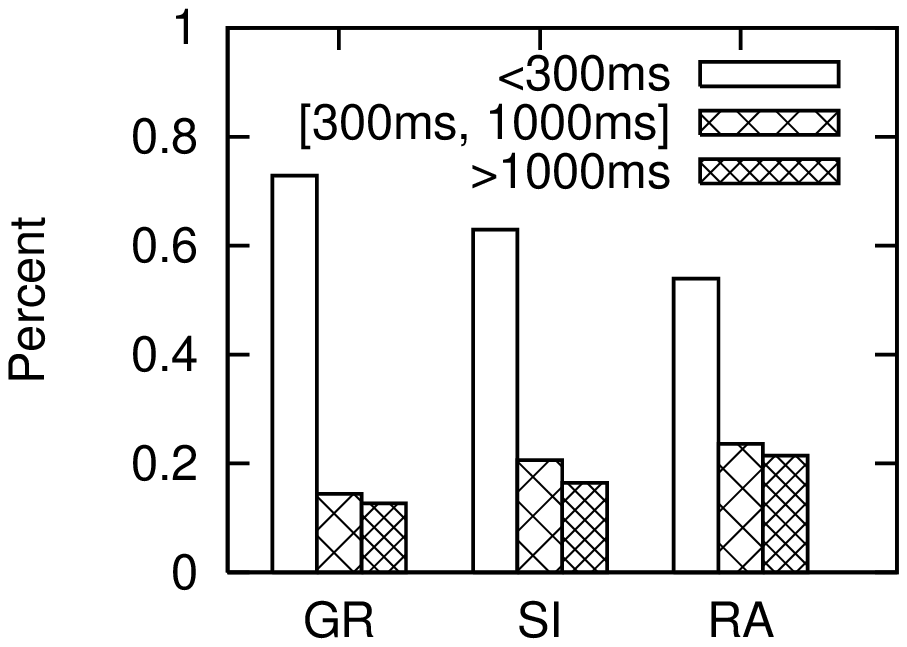}}
\vspace{-1ex}
\caption{Percentage of Latency Values (STS-US-Q1)}
\vspace{-5ex}
\label{fig:migration_percent_latency}
\end{minipage}
\end{figure}

We report the average size of the migrated data and the average time in doing
migration in Figure~\ref{fig:mig_cost_time_1m_us10} and
Figure~\ref{fig:migration_cost_time}.  In each figure, the left y-axis shows
migration cost and the right shows time.  In
Figure~\ref{fig:mig_cost_time_1m_us10}, DP and GR incur the smallest migration
cost and require the least time for migration. DP requires slightly longer
time than GR, since DP needs longer time in selecting the cells to be
migrated.  GR performs the best as shown in
Figure~\ref{fig:migration_cost_time}. In
Figure~\ref{fig:mig_cost_time_5m_us50},
it incurs $30\%$--$40\%$ less migration cost than SI and RA. It takes the
least amount of time in doing migration, which is nearly half of the time
required by RA and $70\%$ of the time required by SI.  Note that the migration
cost (resp.  migration time) in Figure~\ref{fig:mig_cost_time_10m_us50} is
larger than the migration cost (resp.  migration time) in
Figures~\ref{fig:mig_cost_time_5m_us50}.  This is because that a larger number
of \STS queries impose heavier load on the system and the size of each cell to
be migrated becomes larger too.

\paratitle{Comparing latency} To further evaluate the effect of different
algorithms on the system, we also measure the processing latency. The results
are presented in Figure~\ref{fig:mig_percent_1m_us10} and
Figure~\ref{fig:migration_percent_latency}. In
Figure~\ref{fig:mig_percent_1m_us10}, GR has the smallest side effect on the
system, where $80\%$ of tuples are not affected by the migration operations
(less than 100ms) and $4\%$ of tuples are significantly delayed (larger than 1
second). DP is second to GR, where $72\%$ of tuples are not affected by the
migration operations. It results in a larger fraction of tuples being delayed
between 100ms and 1 second. This is due to its long running time in selecting
cells to be migrated.  SI and RA perform worse: SI results in $10\%$
more tuples being delayed than GR. RA results in $20\%$ more tuples being
delayed than GR.

\paratitle{Evaluating the effect of dynamic load adjustments} We evaluate the
benefits of our dynamic load adjustment algorithms by comparing the
throughputs of the system conducting dynamic load adjustments, and the system
without conducting dynamic load adjustments. We use our proposed GR algorithm
in the dynamic load adjustments.  To simulate a workload with varying data
distributions over time, we use query set STS-US-Q3 for this experiment, and
every interval of 10M queries, the types of queries in 10\% of the regions
switch between STS-US-Q1 and STS-US-Q2. We set $\mu=10$M in this experiment. 

\begin{figure}
\centering
\includegraphics[scale=0.6,angle=270]{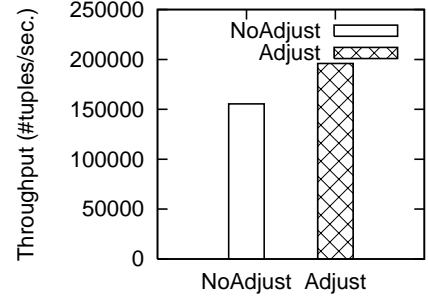}
\caption{The Effect of the Dynamic Load Adjustments}
\vspace{-5ex}
\label{fig:with_without_adjust}
\end{figure}

Figure~\ref{fig:with_without_adjust} shows the experimental results.  The
system with dynamic load adjustments outperforms the system without dynamic
load adjustments by $26\%$ in terms of throughput.  The results demonstrate
the effectiveness of our dynamic load adjustment algorithms.

\section{Conclusion}
\label{sec:conclusion}

In this work, we develop an efficient distributed publish/subscribe system,
   \PSS, over a spatio-textual data stream. We consider the optimal workload
   partitioning problem and propose a new hybrid partitioning
   algorithm.  We also propose effective dynamic load adjustment approaches to
   adjust the load of workers in the scenario of workload changing. Our
   experimental results show that our workload distribution framework performs
   better than the baselines in both throughput and latency, and our dynamic
   load adjustment approaches improve the performance of the system with small
   migration cost.

\section*{Acknowledgement}
\label{sec:ack}

This work was carried out at the Rapid-Rich Object Search(ROSE) Lab at the
Nanyang Technological University, Singapore. The ROSE Lab is supported by the
National Research Foundation, Singapore, under its Interactive Digital
Media(IDM) Strategic Research Programme. This work is also supported in part
by a Tier-1 grant (RG22/15) and a Tier-2 grant (MOE-2016-T2-1-137) awarded by
Ministry of Education Singapore, and a grant awarded by Microsoft. Tom Fu and
Zhenjie Zhang are supported by the research grant for the Human-Centered
Cyber-physical Systems Programme at the Advanced Digital Sciences Center from
Singapore's Agency for Science, Technology and Research (A*STAR). They are
also partially supported by Science and Technology Planning Project of
Guangdong under grant (No. 2015B010131015). We also acknowledge the ``AWS
Cloud Credits for Research'' program for its support of this work.

\small
\bibliographystyle{IEEEtran}
\bibliography{Distributed_Publish-Subscribe_Query_Processing}

\end{document}